\documentclass[
aps,%
10pt,%
final,%
notitlepage,%
oneside,%
onecolumn,%
nobibnotes,%
nofootinbib,%
superscriptaddress,%
nopacs,%
centertags,%
showkeys,%
amssymb,%
]{revtex4} 

\usepackage[utf8]{inputenc}
\usepackage{amsfonts}
\usepackage{amsmath}
\usepackage{amsthm}
\usepackage{amscd}
\usepackage{epsfig}
\usepackage{epstopdf}
\usepackage{graphicx}
  \graphicspath{{/}{figures/}}
  \DeclareGraphicsExtensions{.pdf,.png}
\usepackage[caption=false]{subfig}
\usepackage{todonotes}
\usepackage{tabularx,booktabs}
\usepackage{comment}
\usepackage{enumitem}
\usepackage[pdftex,colorlinks=true]{hyperref}

\newcounter{mycomment}

\SetLabelAlign{parleft}{\parbox[t]{\labelwidth}{\raggedleft\textbf{#1}}}
\setlist[description]{align=parleft,labelindent=0.0ex,labelwidth=7ex}

\newtheorem{Definition}{Definition}

\DeclareMathOperator{\arccosh}{arccosh}
\newcommand{\m}[1]{\ensuremath{\mathrm {#1}}}

\newcommand*{\ee}{\ensuremath{\mathrm{e}}}
\newcommand*{\ii}{\ensuremath{\mathrm{i}}}
\newcommand*{\nn}{\ensuremath{\nonumber}}
\newcommand*{\ICHAIN}{\ensuremath{(\mathrm{ITD}-\mathrm{IMF})\mathrm{chain}}}

\begin{document}

\title{The study of Thai stock market across the 2008 financial crisis}

\author{K. Kanjamapornkul}
\email{kabinsky@hotmail.com}
\affiliation{Department of Computer Engineering, Faculty of Engineering, Chulalongkorn University, Pathumwan, Bangkok 10330, Thailand}
\author{Richard Pinčák}
\email{pincak@saske.sk}
\affiliation{Institute of Experimental Physics, Slovak Academy of Sciences, Watsonova 47, 043 53 Košice, Slovak Republic}
\affiliation{Bogoliubov Laboratory of Theoretical Physics, Joint Institute for Nuclear Research, 141980 Dubna, Moscow Region, Russia}
\author{Erik Bartoš}%
\email{erik.bartos@savba.sk}
\affiliation{Institute of Physics, Slovak Academy of Sciences, Dúbravská cesta 9, 845 11 Bratislava, Slovak Republic
}%

\begin{abstract}
The cohomology theory for financial market can allow us to deform Kolmogorov space of time series data over time period with the explicit definition of eight market states in grand unified theory. The anti-de Sitter space induced from a coupling behavior field among traders in case of a financial market crash acts like gravitational field in financial market spacetime. Under this hybrid mathematical superstructure, we redefine a behavior matrix by using Pauli matrix and modified Wilson loop for time series data. We use it to detect the 2008 financial market crash by using a degree of cohomology group of sphere over tensor field in correlation matrix over all possible dominated stocks underlying Thai SET50 Index Futures. The empirical analysis of financial tensor network was performed with the help of empirical mode decomposition and intrinsic time scale decomposition of correlation matrix and the calculation of closeness centrality of planar graph. 
\end{abstract}

 
\maketitle

\section{Introduction}\label{sec:intro}

Recent studies \cite{stock_stanley2,pincak1,pincak2,pincak3,pincak4} suggest that financial market is a complex dynamical system with underlying nonlinear and nonstationary financial time series data  \cite{stanley4}. When market crashes, the time series data of price of stock will contain with systemic shift and display non-equilibrium entanglement state. The investigation of the 2008 market crash state in nonlinear and nonstationary financial time series data empirically \cite{Stanley,huang}, is one of the  central objectives  in the study of a behavior of traders in financial market and it can be useful model to detect future market crash states. In a spacetime of differential geometry, it is possible to use a cohomology group to deform the Kolmogorov space in time series data to detect a dark state in financial market crash.  Differential geometry of nonstationary and nonlinear predictor and predictant states in mirror symmetry of time series data is one of the active researches. We assume that a nonstationary state can temporally deform the spacetime \cite{wave2} into a homotopic class of supersymmetry breaking of a supersymmetry of dark states \cite{coupling}. The market state is decomposed into a coupling state between two  linear and stationary states \cite{wave3}. One of them is an equivalent of the path of a predictor state of time series of observation. The other is a coincident path of an expectation path of dark state for predictant state of forecasting result evaluation in extradimensions of Kolomogorov space in time series data \cite{kabin1}.

In Walrasian microeconomics the existence of equilibrium paths in the dynamic economic system requires the equilibrium paths of the complete system to coincide with the equilibrium in economics \cite{brouwer} under linearly and stationary assumption of market state space model or dynamic stochastic general equilibrium \cite{dsge} with nonstationary state as a noise in financial market \cite{noise}. Under this assumption, the Brouwer's fixed-point theorem is used to prove the existence of equilibrium price vector, in which it is not known that the algebraic topological structure underlying market is based on an invariant structure of degree over the covering map between cohomology group of sphere \cite{cohomology1}.

The economics is a source of definition of the Walrasian utility function in general equilibrium between supply and demand \cite{neumann} and the topology is concerned with global shape of space and, in particular, its finite or infinite extension \cite{nahm}. The result of interaction of these two theories in the study of time series in the nonstationary state is so called cohomology theory for financial market \cite{cohomo2}. A cohomology theory \cite{cohomology} is a mathematical branch of differential geometry and algebra used to explain anti-de Sitter (AdS) space, it is rich of powerful tools of Cartan calculus of hyperbolic geometry and gauge theory \cite{wave, projective}. 

A general equilibrium price in financial market can be realized as cohomology sequence \cite{cohomology2} of short exact sequences of Kolmogorov space between Riemannian manifold. A generalization of complex plane of equivalent class of supply and demand curve interacts with   Wilson loop of Pauli matrix. It is a spinor field of time series data interpreted as a behavior matrix of traders. Every spinor field of non-orientation state of time series data can be written in the form of quantum triplet state \cite{triplet} in framework of new cohomology theory for financial market.
 
The equilibrium price of data recorded from stock market can provide a source of arbitrage opportunity derived from market nonequilibrium state between supply and demand side. After all behavior traders found the opportunity to gain a profit from financial market, the chance will disappear according to efficient market hypothesis (EMH). The orientation of stock Index Futures market is induced from the orientation of quaternionic field in time series data of underlying stocks, in which we can use the average correlation \cite{stock} as a standard tool for financial network analysis \cite{network,network2}. The ultimate goal of the stock market microstructure prediction \cite{micro} is to find an arbitrage change to detect the general equilibrium point of purchasing power parity \cite{ppp}. It seems that precise definition of a nonstationary state of time series by the projective approach of a hyperbolic space can help to analyze the non-equilibrium state or the crash state in the financial market.

A hyperstructure of non-Euclidean geometry of the behavior of a trader in a financial market \cite{eight} can induce a spin structure of principle  bundle of a correlation matrix by usage of a cocycle over tangent of Kähler manifold. The underlying financial market of such introduced physical quantity of arbitrage opportunity can be realized as a new complicated topological structure related to the curvature of a spacetime of time series data inside an isometry of group action over tangent of higher topological space of market, such as Lie group of predictor and predictant appearing as an evolution feedback path of expectation market state in the Riemanian manifold. It can be generalised to Kähler manifold or Calabi-Yau manifold, with the complex structure of a metric tensor in which it opens the bridge between Yang-Mill theory and the field of expectation induced by traders behaviour to the arbitrage among traders in a financial market. The cohomology sequence of Kolmogorov space of time series data can be induced from a differential 2-form over tangent of Kähler manifold. It is an equivalent class of the path between predictor and predictant topological group. 
The Lie group structure of a correlation matrix induces its tangent manifold as Lie algebra as space of behavior of traders in financial market with a spinor field of double covering space of Kolmogorov space of time series data. This structure can explain the fix point or equilibrium point in financial market network.
In quantum physics, all measurement quantities are associated with Hermitian operator with its eigenvalues. The hermitian property of a correlation matrix induces an asymmetric property of isometry or inertia frame of reference in space of time series data.

The paper is organized as follows. In Section~\ref{sec:theory} we introduce the basic definition of a model of trader behavior and new theoretical construction of de Rahm cohomology for financial market. We also define the another form of cohomology group for physiology of time series data so called knot cohomology in time series data. At the end of this section we define an explicit form of eight market states in Kolmogorov space. In Section~\ref{sec:method} we perform the empirical analysis of tensor correlation using Hilbert-Huang transform from nonstationary and non-orientation state to stationary and orientation state in time series data before sending the result to the planar graph algorithm to build a financial tensor network. The result of tensor network is used to compute the closeness centrality of hyperbolic angle for the detection of a market crash state in time series data. In Section~\ref{sec:results} we show the plots of the results of empirical analysis for daily close price of SET50 Index Futures with average of tensor network of correlation over 42 stocks underlying Index Futures with goal to detect the non-equilibrium state, i.~e., the 2008 financial market crash. We use the empirical analysis method of partial correlation matrix as a main tool. In Section~\ref{sec:discussion} we discuss and make a common conclusion about the result of theoretical derivation of cohomology theory and the empirical result of a detection of 2008 financial market crash with tensor network analysis.

\section{Cohomology Theory in Time Series data}\label{sec:theory}

Let $T^{n}$ be the $n$-dimensional torus with $(S^{1})^{n}=S^{1}\times S^{1}\times \cdots S^{1}$ , the product topology of unit circle into $T^{n}$ a group structure of space time series data. Let $k$-th homology group $H_{k}(T^{n})$ be a free abelian group of rank $C^n_{k}.$ We define the Poincare polynomial of a space of time series data $X_{t}$ as $PX_{t}$.

In classical analysis of time series $x_{t}\in X\subset\mathbb{R}^{n}, X$ is a compact subset, the Poincare polynomial is
\begin{equation}
PX_{t}=1.
\end{equation}
The usage of Poincare polynomial in space of time series data is for measuring the deformation of distance between space of time series and time scale of time series. For classical time series data we then get
\begin{equation}
PX_{t}=1=t_{2}-t_{1}=t_{3}-t_{2}=\cdots.
\end{equation}
If $X_{t}=S^{1}$ we get $PX_{t}=x$, that means
\begin{equation}
PX_{t}=1+x,
\end{equation}
then
\[t_{2}-t_{1}=1+(t_{2}-t_{1}).\]
The distance in time series on sphere is not constant. But it is an equivalent class of complex projective space by usage of stereoprojection from real line to unit sphere, where $\infty$ is send to $1$ in north pole in the sense of the context of Poincare polynomial of time series data.
There does not exist a cohomology group of space of time series data in which we can deform space over time period in time ordering like time series data. Our main goal is to create such cohomology theory.

Let $X_{t}$ be a Kolmogorov space of time series data. There exists the relationship between homotopy class $[S^{n},X_{t}]$ and cohomology group $H^{n}(X_{t})$ in algebraic topology
\begin{equation}
[S^{n},X_{t}]=H^{n}(X_{t}):=H^{n}(x_{t};[G,G^{\ast}])
\end{equation}
and in this case we set $[G,G^{\ast}]=\mathbb{Z}/2$ for 2 states of entanglement and non-entanglement state of prediction. 
For given financial time series data, $x_{1}\rightarrow x_{2}\rightarrow x_{3}\rightarrow\cdots x_{n}$, we induce a cohomology theory for financial market for time series data in which we can measure the invariance property of space of time series over time. We use the notation for this new type of cohomology of time as $HHH^{n}_{t}(X)$
\begin{gather}
HHH_{t}^{n}(X_{t}):  H^{n}(x_{1};[G,G^{\ast}]) \rightarrow  H^{n}(x_{2};[G,G^{\ast}])  \rightarrow  H^{n}(x_{3};[G,G^{\ast}])\rightarrow \cdots  \rightarrow H^{n}(x_{t};[G,G^{\ast}]),\nn\\
HHH_{t-1}^{n-1}(X_{t}):  H^{n-1}(x_{1};[G,G^{\ast}]) \rightarrow  H^{n-1}(x_{2};[G,G^{\ast}])  \rightarrow  H^{n-1}(x_{3};[G,G^{\ast}])\rightarrow \cdots  \rightarrow H^{n-1}(x_{t-1};[G,G^{\ast}]),\nn\\
\cdots\\
HHH_{2}^{2}(X_{t}):  H^{2}(x_{1};[G,G^{\ast}]) \rightarrow  H^{2}(x_{2};[G,G^{\ast}])  \nn\\
\cdots\nn
\end{gather}
with
\begin{equation}
0\rightarrow HHH_{t}^{n}(X_{t}) \rightarrow  HHH_{t-1}^{n-1}(X_{t})  \rightarrow  \cdots\rightarrow \cdots  \rightarrow HHH_{0}^{0}(X_{t})\rightarrow 0.
\end{equation}
The cohomology in this context is used to measure the high dimensional invariance property of physiology of space time series data at the end point of time series measurement.
 
\begin{Definition}
Let $G$ be a Lie group of predictor, the path of result of prediction. Let $G^{\ast}$ be a Lie group of predictant, the path of evaluation of result of prediction without of sample data. A time series $x_{t}$ is a path connected evolution feedback component with correct expectation path if and only if for every $n,t>0$
\begin{equation}
HHH^{n}_{t}(X_{t})=HHH^{n}_{t}(x_{t};[G,G^{\ast}])=\{[0]\}
\end{equation}
for every  $n$-dimension of sphere to measure the invariant property of space over time scale.
\end{Definition}
A spin group is a principle bundle with double covering and a following sequence is well known
\begin{equation}
1\rightarrow \mathbb{Z}/2\rightarrow \m{Spin}(2)=\m{U}(1)\rightarrow \m{SO}(2)\rightarrow 1 \label{spin1}
\end{equation}
in which we can induce infinite sequence of cohomology group.
\begin{Definition}
An equivalent class of up $[up(x_{t})]\in \mathbb{Z}/2$ and down $[down(x_{t})]\in \mathbb{Z}/2$ orientation state of time series data is an equivalent class of $\m{Spin}(2)$ over $\m{SO}(2)/\m{Spin}(2)=\mathbb{Z}/2$ in short exact sequence in Eq.~(\ref{spin1}). 
\end{Definition}

The exact sequence Eq.~(\ref{spin1}) can be applied to every dimensions, e.~g., for $n=3$ we have a high dimensional sphere $S^{7}$ with Hopf fibration as a space underlying hidden dynamical system of financial market. If we have two cutting lines $L_{1}$, $L_{2}$ not perpendicular to each other in $S^{7}$, where $S^{7}/\m{Spin(3)}\simeq \mathbb{H}P^{1}$, the supply $S=L_{1}(x,a_{1},b_{1})\in \mathbb{H}$ and demand  $D=L_{2}(y,a_{2},b_{2})\in \mathbb{H}$ of financial market intersects each other. This two lines will twist across each other in equilibrium state like Möbius operator in Markov trace. In equilibrium state we can transfer  the measurement from one line to another line by using projective geometry, simply by interchange of coordinate system of supply $S$ to demand side $D$  of financial market by
\begin{equation}
|\mathcal{S}(p)|=|\mathcal{D}(p)|
\end{equation}
where $p$ is a price. Let $D=\ln |\mathcal{D}|$ and $S=\ln |\mathcal{S}|$ so we have $\ln |\mathcal{S}(p)|=\ln |\mathcal{D}(p)|$. If $D=(x_{1},x_{2},x_{3},\cdots)$ and $S=(y_{1},y_{2},y_{3},\cdots) $ we induce a projective coordinate
\begin{equation}
z=\frac{S}{D}=\frac{x}{y}=\frac{\lambda x_{1}}{\lambda y_{1}}\stackrel{\m{Twistor}}{\longrightarrow}\frac{y}{x}=\frac{D}{S}=\frac{1}{z}.
\end{equation}

\begin{Definition}
A Wilson loop of time series data, denoted as $W_{x_{t}}(\sigma_{i})$ is a twistor in complex projective space $\mathbb{C}P^{1}$. It turn one side of Euclidean plane twist into another side behind the plane with a help of modified Möbius map $z=\frac{1}{z}$. 
\end{Definition}
Therefore, we have $W_{x_{t}}(\frac{\lambda S}{\lambda D})=\frac{D}{S}$. The equivalent class of supply and demand $S\in [s]$, $D\in [d]$ is induced from the gluing process of $0$ to $\infty$ by modified Möbius map $z=\frac{1}{z}$, such that $[s] \sim [d]$ if and only if there exist $\lambda_{i}$ where $x_{i}=\lambda_{i} y_{i}$. The projective coordinate is a local coordinate on Minkowski space. Let $g_{ij}$ be Jacobian matrix of supply and demand, we have a shallow to hidden state of transformation by using Wigner ray transform
\begin{equation}
g_{ij}\mapsto \lambda g_{ij},\,\lambda \in \mathbb{U}(1)\simeq \m{Spin}(2)
\end{equation}
where $\lambda =\sigma_{x}$ and $\sigma_{x}$ are Pauli matrices in $\m{Spin}(2)$ group. 

\begin{Definition}\label{def:w}
Let $S$ be supply linear compact operator in Banach space and $D$ be demand operator. A Wigner ray of time series data at general equilibrium point in financial time series data is a point of price $x_{t}(S,D)$ such that there exist a ray of unitary operator $\lambda \in \m{SU}(2)\simeq \m{Spin}(3)$
\begin{equation}
W_{x_{t}}<S,D>:= <\lambda S,\lambda D>=\lambda<S,D>=<S,D>.
\end{equation}
\end{Definition}

\begin{figure}[!t]
	\centering
	\subfloat[]{
		\includegraphics[height=.3\textheight]{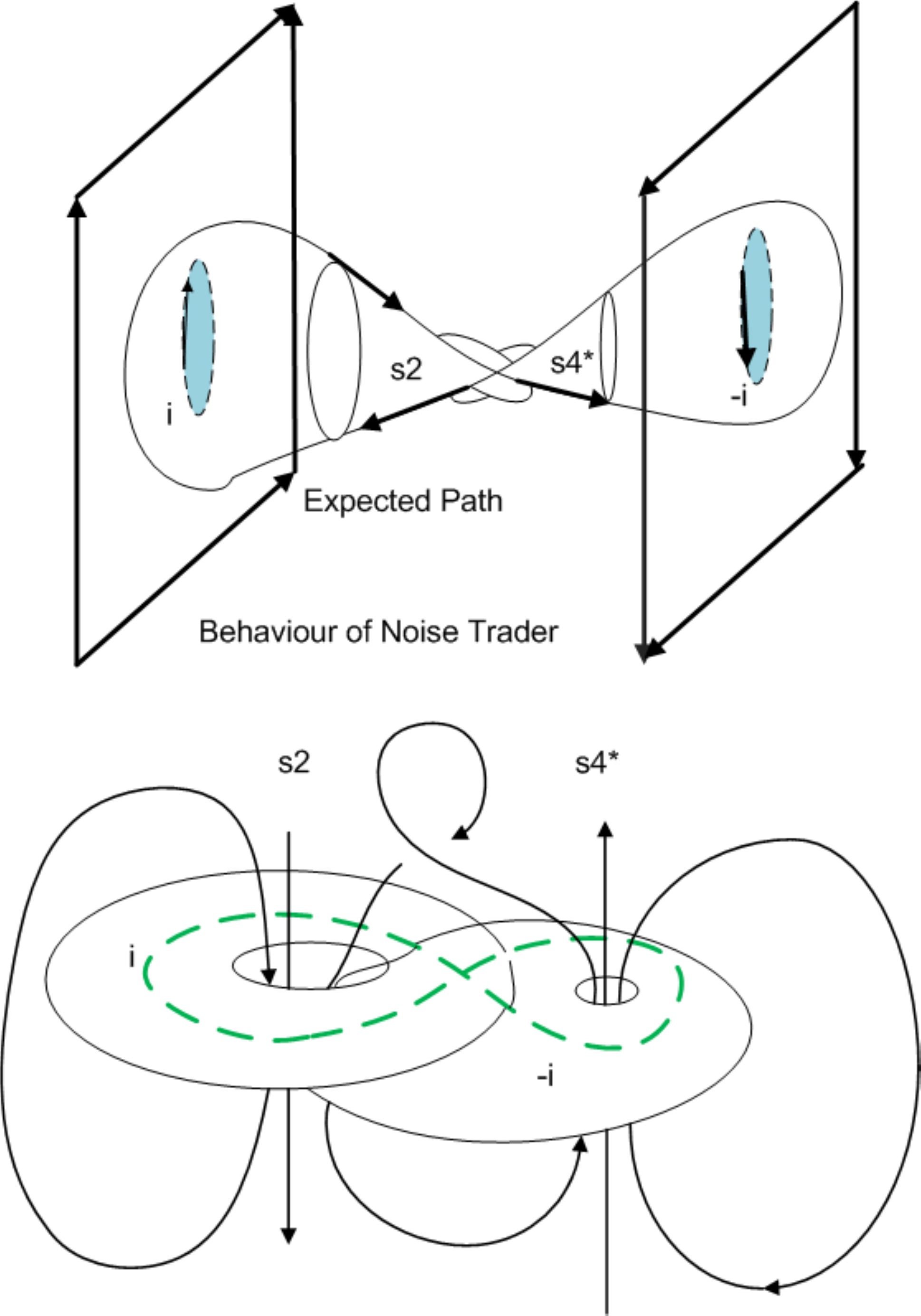}\label{knot}
	}\hspace{5em}
	\subfloat[]{
		\includegraphics[height=.3\textheight]{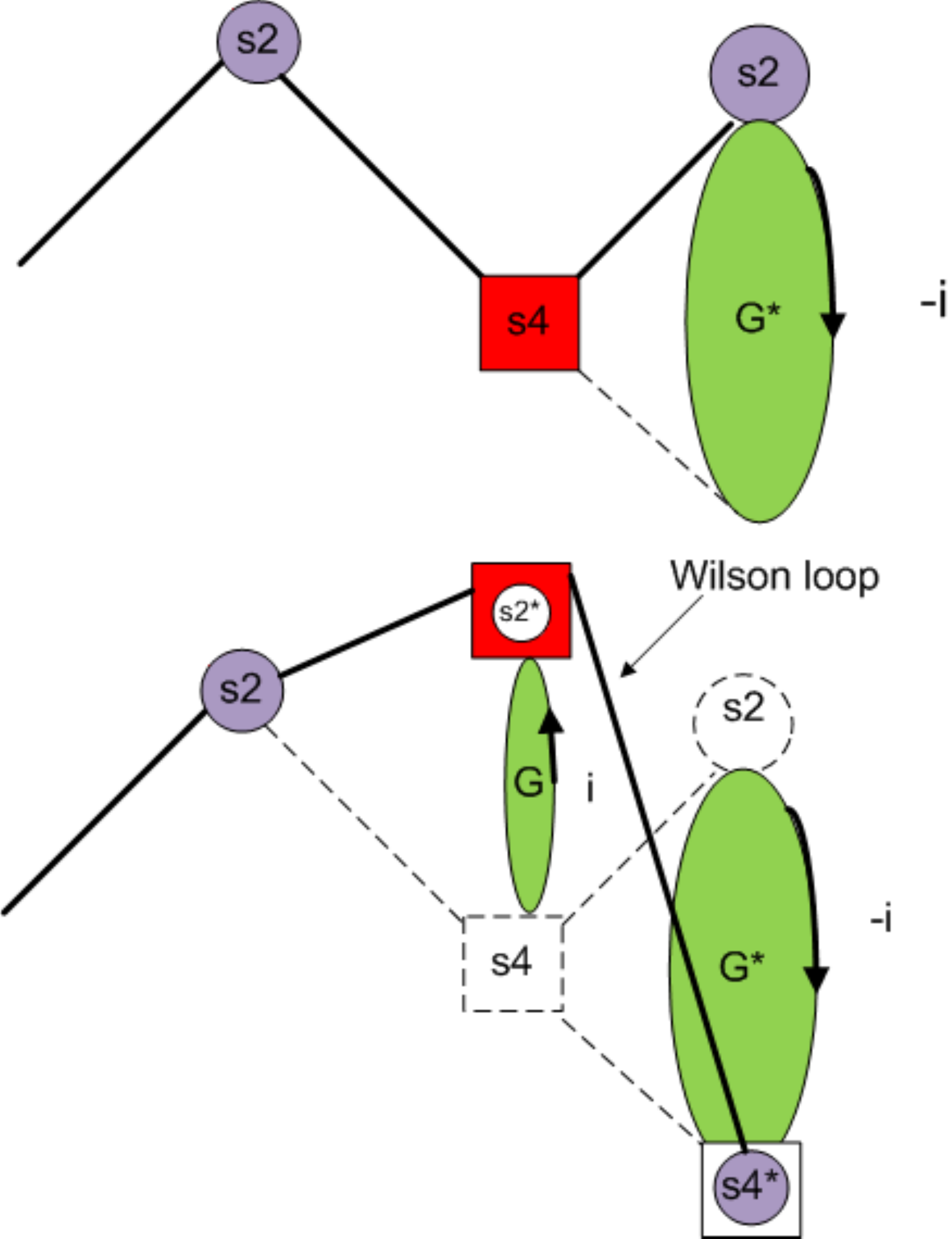}\label{wilson}
	}
	\caption{(a) The knot model of time series data in knot cohomology model between D-brane and anti-D-brane of time series data, (b) Wilson loop of time series data in the form of entanglement state coupling between predictor and predictant states in end point of time series data.}
\end{figure}

We introduce three types of Killing vector field for market potential field of behavior of traders in market.
\begin{gather}
A_{\pm,1}(f)=\frac{\partial}{\partial f}: \m{SO}(2)\rightarrow \m{Spin}(2):M\rightarrow T_{x}M \nn\\
A_{\pm,2}(\sigma)=\frac{\partial}{\partial \sigma}: \m{SO}(2)\rightarrow \m{Spin}(2):M\rightarrow T_{x}M\\
A_{\pm,3}( \omega)=\frac{\partial}{\partial \omega}:\m{SO}(2)\rightarrow \m{Spin}(2):M\rightarrow T_{x}M\nn
\end{gather}
where ``$+$'' means optimistic behavior of trader, ``$-$'' means pessimistic trader, $f$ is a fundamentalist trader, $\sigma$ is a noise trader and $\omega$ is a bias trader. $A_{1}$ is an agent field of adult behavior of trader as fundamentalist, $A_{2}$ is an agent field of teenager behavior of trader as herding behavior or noise trader, $A_{3}$ is an agent field of child behavior of trader or bias trader. The behavior states are spin up or down as the expected states in market communication layer of transactional analysis framework \cite{eric2}. The strategy of fundamentalist trader is to expect price in which period price is in minimum point $s_{4}$ and maximum point $s_{2}$ (crash and bubble price). Then the optimistic fundamentalist $f_{+}$ will buy at minimum point $s_{4}$ of price below fundamental value and the pessimistic fundamentalist $f_{-}$ will sell product or short position at maximum point of price if the maximum point is over a fundamental value. The strategy of noise trader $\sigma_{\pm}$ is different from fundamentalist. The up state is signified as optimistic market expected state and down state as pessimistic as market crash state of intuition state of forward looking trader \cite{eric4}.
An equivalent class of supply induced from interaction of behavior of fundamentalist and noise trader in mirror symmetry model of expected on arbitrage opportunity over physiology of time series data in entanglement states between $s_{2}$, $s_{4}$ and $s_{2}^{\ast}$, $s_{4}^{\ast}$ is shown in Fig.~\ref{knot} (the definition of $s_{2}$, $s_{4}$, $s_{2}^{\ast}$ and $s_{4}^{\ast}$ can be found in \cite{kabin1}).

We can use Pauli matrix $\sigma_{i}(t)$ to define a strategy of all agents in stock market as basis of quaternionic field span by noise trader and fundamentalist. Let $+1$ stands for a buy at once and $-1$ for a sell suddenly. Let $+i$ be inert to buy and $-i$ be inert to sell. The row of Pauli matrix represent the position of predictor and the position in column represent the predictant state.

\begin{Definition}
For noise trader, we use finite state machine for physiology of time series to accepted pattern defined by
\begin{equation}
A_{2}:=\sigma(t)= \sigma_{y} =  \left[   \begin{array}{ccc}
 &s_{2}{\ast}(t) &s_{4}^{\ast}(t)\\
s_{2}(t)&0&-i\\
s_{4}(t)&i&0\\
\end{array}\right ]=  \left[   \begin{array}{ccc}
 &s_{2}{\ast}(t) &s_{4}^{\ast}(t)\\
s_{2}(t)&0&\sigma_{+}\\
s_{4}(t)&\sigma_{-}&0\\
\end{array}\right ]
\end{equation}
where $\sigma_{y}$ is a Pauli spin matrix.
\end{Definition}

\begin{Definition}
For fundamentalist trader, we use finite state machine for physiology of time series to accepted pattern defined by
\begin{equation}
 A_{1}:=f_{\pm}=-W(\sigma_{z}) =  \left[   \begin{array}{ccc}
 &s_{2}^{\ast} &s_{4}^{\ast}\\
s_{2}(t)&0&-1\\
s_{4}(t)&1&0\\
\end{array}\right ]=\left[   \begin{array}{ccc}
 &s_{2}^{\ast} &s_{4}^{\ast}\\
s_{2}(t)&0&f_{-}\\
s_{4}(t)&f_{+}&0\\
\end{array}\right ]
\end{equation}
\end{Definition}

Let $\omega$ be a bias behavior $A_{3}$ of market micropotential field. From $[\sigma_{y},\sigma_{z}]=2i\sigma_{x}$ follows 
\begin{equation}
[f,-W^{-1}(\sigma)]=2i \omega,
\end{equation}
where $W$ is a Wilson loop (knot state between predictor and predictant) for time series data and $W^{-1}$ is inverse of Wilson loop (unknot state between predictor and predictant). We have an entanglement state (Fig.~\ref{wilson}) induced from strategy of noise trader as herding behavior explained by
\begin{equation}
s_{2} \stackrel{-i}{\rightarrow } s_{4}^{\ast},\quad s_{4} \stackrel{i}{\rightarrow } s_{2}^{\ast}
\end{equation}
with
\begin{equation}
W_{x_{t}}\bigg( \left[   \begin{array}{c}
 s_{2}  \\
s_{4}^{\ast} \\
\end{array}\right ] \bigg)= \left[   \begin{array}{c}
 s_{4}^{\ast}  \\
s_{2} \\
\end{array}\right ] 
\end{equation}

\subsection{de Rahm coholomogy for financial market}

In this section we explain the source of second cohomology group of spinor field in financial time series data. It comes from the entanglement state of field of behavior of trader $A_{i}$ with respect to an equivalent class of error of expectation of out coming of physiology of time series data $\partial[s_{2}]=[s_{2}]^{\ast}-[s_{2}]$ and $\partial[s_{4}]=[s_{4}]^{\ast}-[s_{4}]$.

The mathematical structure in this section is related to Chern-Simon theory so called gravitational field in three forms of connection $F^{\bigtriangledown}\in \Omega^{2}(T_{x}M\otimes T_{x}^{\ast}M)$ over three vector fields in financial market, $\Psi_{i}\in M$ a market state field, $\mathcal{A}_{i}$ is an agent behavior field and $[s_{i}]$ is a field of physiology of financial time series data over Kolmogorov space.   Let $\mathcal{A}=(A_{1},A_{2},A_{3})$ be market micro potential field of agent or behavior of trader in market. Let $\Psi_{i}^{(\mathcal{S},\mathcal{D})}([s_{i}],[s_{j}])$ be a field of market $8$ states of equivalent class of physiology of time series $[s_{i}],[s_{j}]$. 

\begin{figure}[!t]
\centering
\includegraphics[width=.55\textwidth]{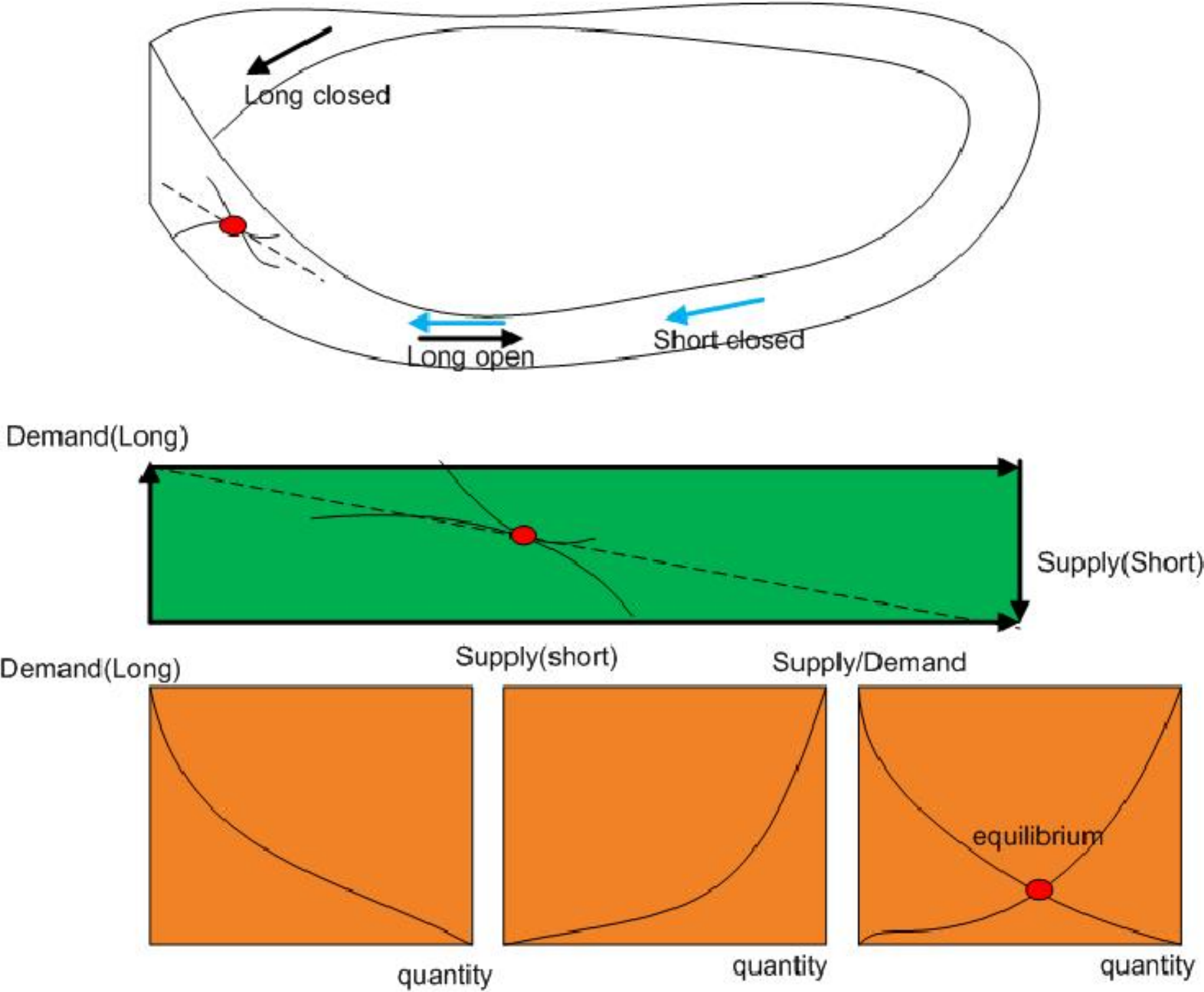}
\caption{The order submission as time series data in complex projective space.\label{twist}}
\end{figure}

In financial market microstructure \cite{micro2} we define a new quantity in microeconomics induced from the interaction of order submission from  supply and demand side of the orderbook, so called market micro  vector potential \cite{micro} twist field (Fig.~\ref{twist}) or connection in Chern-Simon theory for financial time series $\mathcal{A}(x,t;g_{ij},\Psi_{i}^{(\vec{\mathcal{S}},\vec{\mathcal{D}})},s_{i}(x_{t}))$, where  $\Psi_{i}^{(\vec{\mathcal{S}},\vec{\mathcal{D}})}$ is a state function for supply $\vec{\mathcal{S}}$ and demand  $\vec{\mathcal{D}}$ of financial market. Let $ \Psi_{i}^{(\vec{\mathcal{S}},\vec{\mathcal{D}})}(F_{i})$ be a scalar field induced from the news of market factor $F_{i}$, $i=1,2,3,4$. Let $x_{t}(\mathcal{A},x^{\ast},t^{\ast})$ be a financial time series induced from the field of market behavior of trader, where $x^{\ast}$ and $t^{\ast}$ are hidden dimensions of time series data in Kolmogorov space. 

\begin{Definition}
Let $S$ be a supply  potential field defined by rate of change of supply side of market potential fields of behaviour trader in induced field of behavior of agent $\mathcal{A}$ in stock market.
\begin{align}
\vec{\mathcal{S}}&=-\frac{\partial \mathcal{A}}{\partial [s_{4}]},\quad
\vec{\mathcal{NS}}=-i\frac{\partial \mathcal{A}}{\partial [s_{4}]},\quad
\vec{\mathcal{D}}=\frac{\partial^{2} \mathcal{A}}{\partial^{2} [s_{2}]}\quad
\vec{\mathcal{ND}}=+i\frac{\partial^{2} \mathcal{A}}{\partial^{2} [s_{2}]}
\end{align}
with
\begin{equation}
\mathcal{A}=(A_{1}(f),A_{2}(\sigma),A_{3}(\omega)).
\end{equation}
\end{Definition}
 
\begin{Definition}
Let $\vec{\mathcal{D}}$ be a demand  potential field (analogy to magentic field). It is
 defined by rate of change of market potential fields in induced field of $\vec{\mathcal{S}}$,
\begin{equation}
\vec{\mathcal{D}}=\bigtriangledown \times \vec{\mathcal{S}}.
\end{equation}
\end{Definition}

\begin{Definition}
Let $F_{\mu\nu}$ be a stress tensor for market microstructure with component $A_{\mu}$ of market potential field (the analogy with connection one form in Chern-Simon theory) defined by the rank 2 antisymmetric tensor field strength
\begin{equation}
F_{\mu\nu} =\partial_{\mu}A_{\nu}-\partial_{\nu}A_{\mu},\quad
\frac{d\mathcal{A}}{d[s_{2}]}=S_{\nu}=F_{\mu\nu}\Psi^{\nu}, \hspace{0.5cm}  \frac{d\mathcal{A}}{d[s_{4}]}=  D_{\mu}=\epsilon_{\mu\nu}^{\kappa\lambda}F_{\kappa\lambda}\Psi^{\nu}.
\end{equation}
\end{Definition}
These three vector fields of a behavior of trader play role of 3-form in tangent of complex manifold. It is an element of section $\Gamma$ of manifold of market $\mathcal{A}\in \Gamma(\wedge^{3}T^{\ast}_{x}M\otimes \mathbb{H})= \Omega^{3}(M).$

\begin{figure}[!t]
\centering
\includegraphics[width=.45\textwidth]{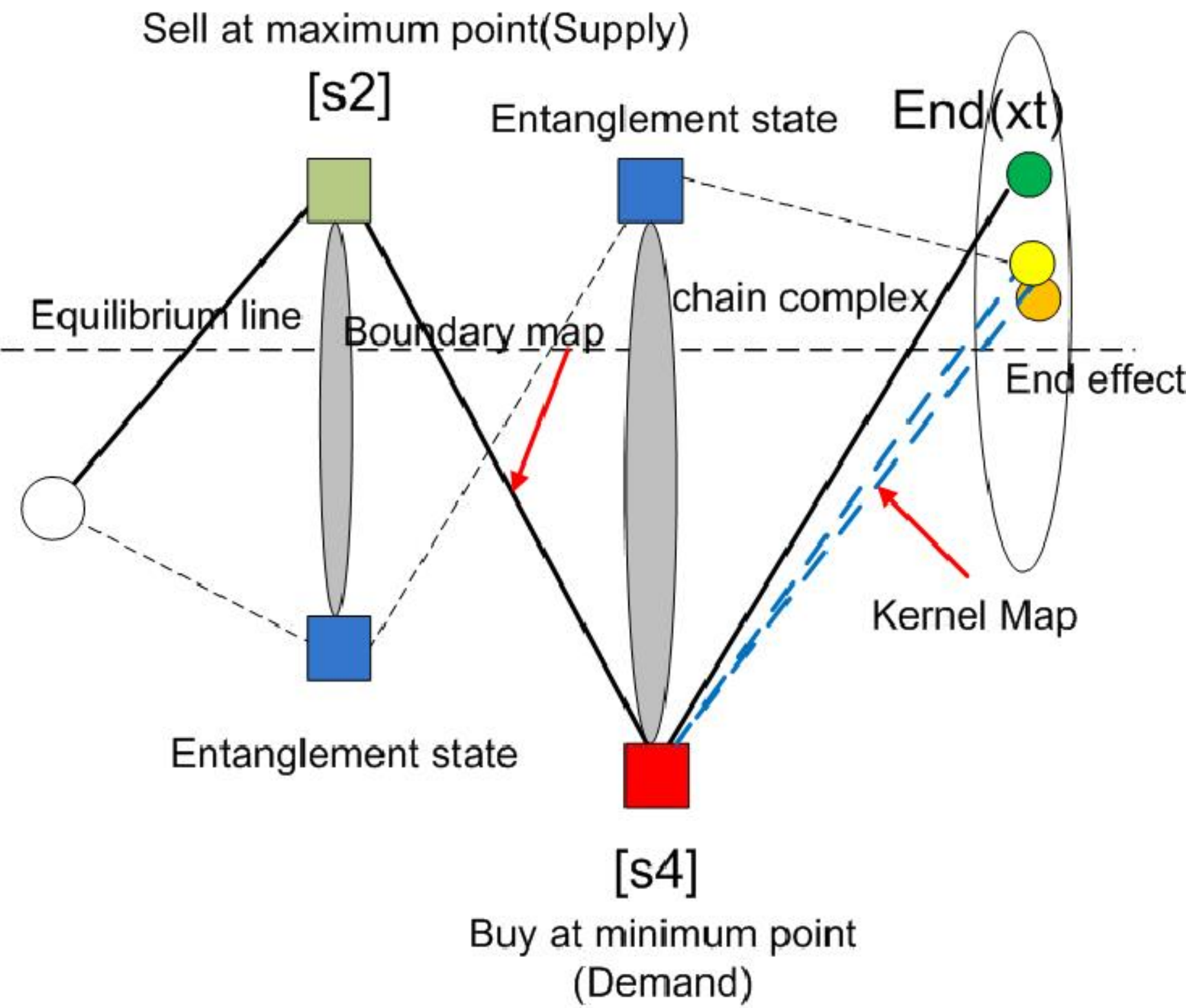}
\caption{The cohomology map between predictor and predictant states\label{cohomlogy}.}
\end{figure}

Let consider 3-differential form of behavior of trader as market potential field $\mathcal{A}$
\begin{gather}
d\mathcal{A}=\sum_{ijk=1,2,3}F_{ijk}^{ \bigtriangledown_{[s_{i}]}}dA_{i}\wedge dA_{j}\wedge dA_{k} 
\\
\begin{CD}
\cdots @<<<\Omega^{3}(M) @<{\text{d}}<< \Omega^{2}(M) @<{\text{d}}<< \Omega^{1}(M)@<{\text{d}}<< \Omega^{0}(M)@<{\text{d}}<< 0 \\
    @.   @VVV @VVV @VVV @VVV @.\\
\cdots @<<<C_{3}(X_{t}) @>\partial>> C_{2}(X_{t}) @>\partial>> C_{1}(X_{t})@>\partial>> C_{0}(X_{t})@>\partial>> 0
\end{CD}
\end{gather}
with a market state $\Psi^{(S,D)}\in C^{1}(M)$.
de Rahm cohomology for financial market is defined by using the equivalent class over this differential form. The first class is so called market cocycle $Z^{n}(M)=\{f^{\ast}: C^{n}(M)\rightarrow C^{n+1}(M)  \}$ with commutative diagram between covariant and contravariant functor
\begin{equation}
\begin{CD}
\cdots @<<<H^{3}(M) @<{\text{d}}<< H^{2}(M) @<{\text{d}}<< H^{1}(M) @<{\text{d}}<< H^{0}(M)@<{\text{d}}<<0\\
@.  @VV\xi V @VV\xi V @VV\xi V @VV\xi V @VV\xi V\\
\cdots @<<<H_{3}(X_{t}) @>\partial>> H_{2}(X_{t}) @>   \partial>> H_{1}(X_{t})@>   \partial>> H_{0}(X_{t})@>   \partial>> 0\\
@. @VV\nu V @VV\nu V @VV\nu V @VV\nu V @VV\nu V\\
\cdots @<<<H^{3}(x_{t}:[G,G^{\ast}]) @<{\text{d}}<< H^{2}(x_{t}:[G,G^{\ast}]) @<{\text{d}}<< H^{1}(x_{t}:[G,G^{\ast}]) @<{\text{d}}<< H^{0}(x_{t}:[G,G^{\ast}])@<{\text{d}}<<0\\
\end{CD}
\end{equation}
It is a kernel of co-differential map between supply and demand to market potential field
\begin{equation}
\m{Ker}(d:\Omega^{2}(M)\rightarrow \Omega^{3}(M)), d:[s_{i}](S,D)\mapsto \mathcal{A}(A_{1},A_{2},A_{3}),d[s_{i}](S,D)=0=\mathcal{A},
\end{equation}
where $[s_{i}](S,D)$ is an equivalent class of physiology of time series data. The boundary map of cochain of market $\Omega^{2}(M)$ is a second equivalent class used for modulo state,
\begin{equation}
\m{Im}(d:\Omega^{1}(M)\rightarrow \Omega^{2}(M), d:\Psi^{(S,D)}\mapsto [s_{i}](S,D).
\end{equation}

\begin{figure}[!t]
\centering
\includegraphics[width=.3\textwidth]{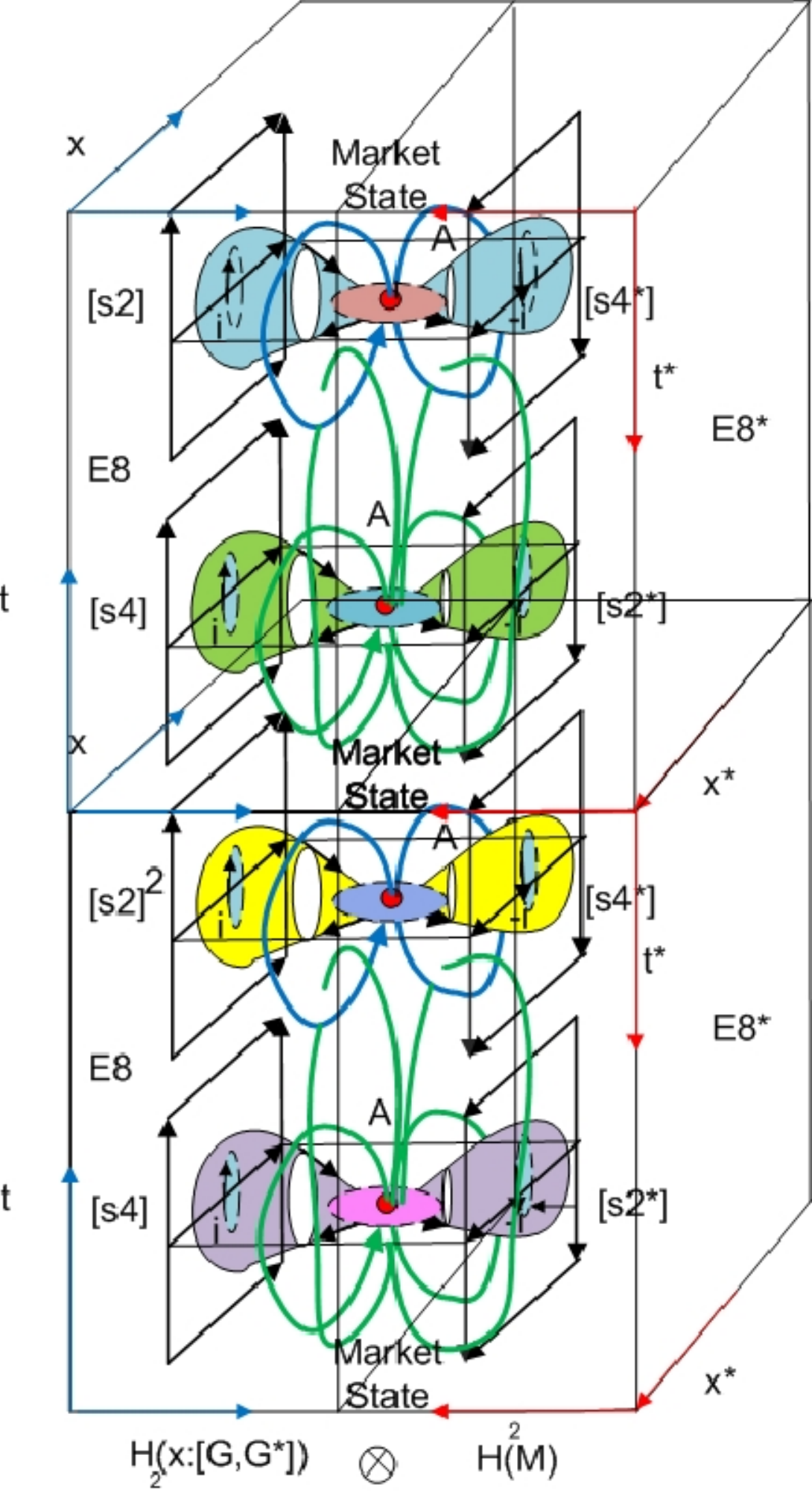}
\caption{Figure shows D-brane and anti D-brane of Calabi-Yau hypersurface of eight market states in time series model. The Euclidean plane cannot visualize the physiology of market state in this model because all  of eight states lay in the extradimensions of Kolmogorov space in time series data\label{plane}.}
\end{figure}

\begin{Definition}
The de Rahm cohomology for financial time series is an equivalent class of second cochain of market (Fig.~ \ref{plane})
\begin{equation}
H_{DR}^{2}(M)=\m{Ker}(d:\Omega^{2}(M)\rightarrow \Omega^{3}(M))/\m{Im}(d:\Omega^{1}(M)\rightarrow \Omega^{2}(M)).
\end{equation}
\end{Definition}
The meaning of a new defined mathematical object is in use in measurement of a market equilibrium in algebraic topology approach by AdS Yang-Mills field in finance. From the definition above we get a communication behavior between two sides of market by the twist field of behavior trader of differential 3-form in financial market with underlying Kolmogorov space $X_{t}$ by
\begin{equation}
  *F^{ \bigtriangledown_{[s_{i}]}}=\oint _{H^{2}_{DR}(M)}\Psi_{k}^{(\mathcal{S},\mathcal{D} )}F_{\mu\nu}^{k}A_{\nu} =F^{ \bigtriangledown_{[s_{i}^{\ast}]}}.
\end{equation}
The section of manifold induces a connection of differential form. The connection is used typically for gravitational field, in econophysics, the connection is used for measurement of arbitrage opportunity, we have
\begin{equation}
\sum_{ij}g_{ij}\frac{\partial A_{i}}{\partial [s_{2}]} \wedge \frac{\partial A_{j} }{\partial [s_{4}]} \in H_{DR}^{2}(X_{t}). 
\end{equation}
 
We simplify the model of financial market in unified theory of $E_{8}\times E_{8}^{\ast}$ (Fig.~\ref{plane}) under mathematical structure of Kolmogorov space underlying time series data. Let $M\in\mathbb{H}P^{1}/\m{Spin}(3)$ be the market with supply and demand sides and $M^{\ast}:\rightarrow\mathbb{H}P^{1}/\m{Spin}(3)\rightarrow \mathbb{H}P^{1}$ be dual market of behavior of noise trader $\sigma$ and fundamentalist $f$. Then we prove an existence of general equilibrium point in the market by using the equivalent classes of supply $[s]$ and demand $[d]$ in the complex projective space $[s],[d]\in \mathbb{C}P^{1}$. 
Kolmogorov space $S^{7}=\mathbb{H}P^{1}\simeq E_{8}$ can be used as explicit state of financial market and its dual $\mathbb{H}P^{1\ast}/spin(3)\simeq E_{8}^{\ast}$ as dark state in financial market. It is a ground field of first knot cohomology group tensor with first de Rahm cohomology group $H_{1}(x_{t};[G,G^{\ast}];\mathbb{H}P^{1}/\m{Spin}(3))\otimes H^{1}(M;\mathbb{H}P^{1\ast}/\m{Spin}(3))$. We induce a free duality maps $\xi:H_{1}(x_{t};[G,G^{\ast}];\mathbb{H}P^{1}/\m{Spin}(3))\rightarrow H^{1}(M;\mathbb{H}P^{1\ast}/\m{Spin}(3)) $ as market state $E_{8}\times E_{8}^{\ast}$ in grand unified theory model of financial market.
For simplicity we let a piece of surface cut out from Riemann sphere with constant curvature $g_{ij}$ define as a parameter of D-brane field basis with the same orientation   as a defintion of supply and demand in complex plane (flat Riemann surface). Now we blend complex plane to hyperbolic coordinate in eight hidden dimensions of $\Psi=(\Psi_{1},\cdots \Psi_{8}^{\ast})\in S^{7}$.

\begin{figure}[!t]
	\centering
	\subfloat[]{
		\includegraphics[width=.21\textheight]{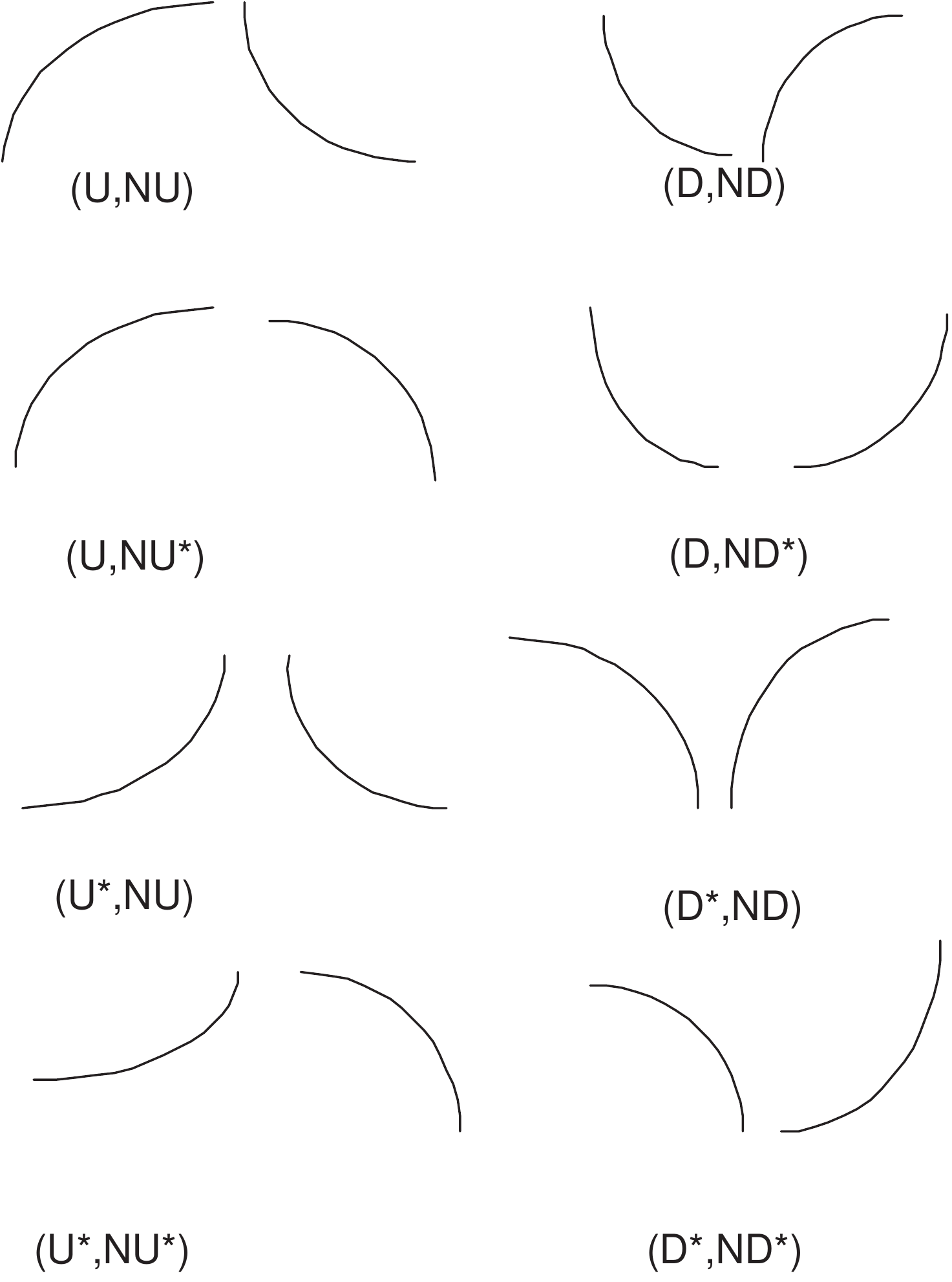}\label{state_agent}
	}\hspace{3em}
	\subfloat[]{
		\includegraphics[width=.25\textheight]{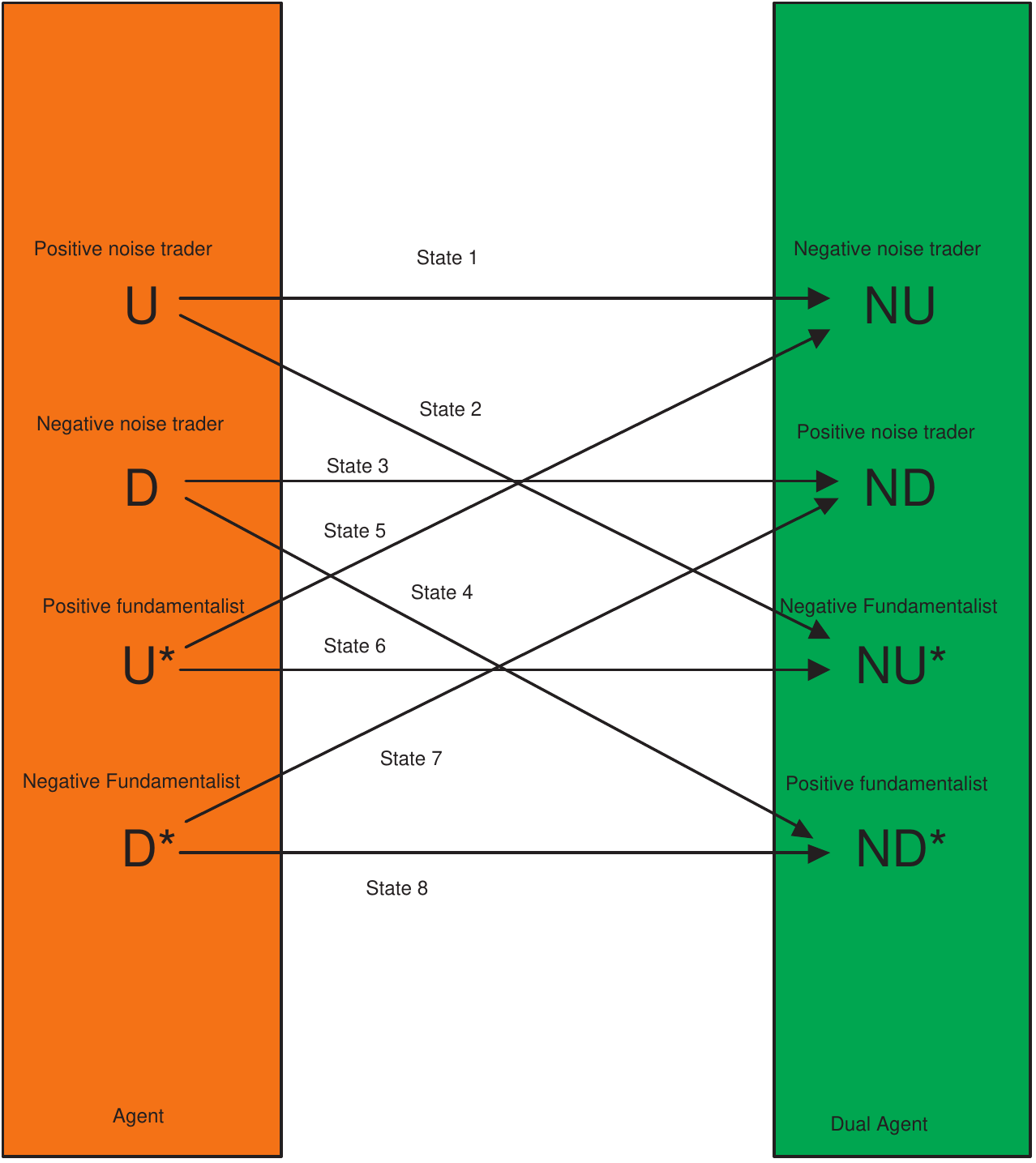}\label{market_state}
	}
	\caption{(a) Picture shown type I of market field. It is originally denoted as market 8 states. The first block in this diagram is predictor Lie group $G$. The second block is predictant Lie group $G^{\ast}$ as dark state. 
		A market field is an  evolution feedback path communicate between these 2 manifolds (b) Picture shown a physiology of shape time series data in complex plane of our definition of 8 states of coupling between behavior of traders in financial market .}
\end{figure}

\begin{Definition} 
Let $[G,G^{\ast}]$ be homotopy class between predictor and predictant Lie group, $G\subset S^{1}$ and $G^{\ast}\subset S^{1}$ (Fig.~\ref{state_agent}). The physiology shape of eight states is shown in Fig.~\ref{market_state}.
We assume that  the ground state of market potential field basis  over hidden state of time series data (not D-brane $E_{8}\times E_{8}^{\ast}$, just projection of boundary of D-brane).  Let $\mathbb{C}P^{1}\simeq [G,G^{\ast}]\ni \Psi_{i}$, $i=1,\dots, 8$ defined by
\begin{equation}
\Psi_{i}^{(\vec{\mathcal{S}},  \vec{\mathcal{D}})}\in [G,G^{\ast}]\subset\mathbb{C}P^{1}\simeq S^{2}  \simeq 
S^{3}/S^{1}\rightarrow S^{3}\rightarrow S^{7}\rightarrow \mathbb{H}P^{1}.
\end{equation}

\textbf{\em State 1.} induced  from coupling  between supply side of market and supply side in dual market 
\begin{equation}
 \Psi_{1}^{(\mathcal{S},\mathcal{D})}([s_{i}],[s_{j}]^{\ast}) =(S,NS)\stackrel{\subset}{\rightarrow}([U],[NU])= (\Psi_{1}(\sigma_{+}), \Psi_{1}^{ \ast}(\sigma_{+}))=([\ee^{\ii\theta }],[\ee^{\ii(-2\pi+\gamma)}]) ,0<\theta<\frac{\pi}{2},0<\gamma<\frac{\pi}{2} \nonumber
\end{equation}

\textbf{\em State 2.} induced from coupling between   
\begin{equation}   
\Psi_{2}^{(\vec{\mathcal{S}},\vec{\mathcal{D}})}([s_{i}],[s_{j}]^{\ast})=(S,NS^{\ast})\stackrel{\subset}{\rightarrow} ([U],[NU^{\ast}])= (\Psi_{2}(\sigma_{+}), \Psi_{2}^{ \ast}(f_{-}))=([\ee^{\ii\theta }],[\ee^{\ii(-2\pi+\gamma )}]) ,0<\theta<\frac{\pi}{2},\frac{3\pi}{2}<\gamma<2\pi
\nonumber
\end{equation}

\textbf{\em State 3.} induced from coupling between 
\begin{equation}                                                                  
\Psi_{3}^{( \vec{\mathcal{S}},\vec{\mathcal{D}})}([s_{i}],[s_{j}]^{\ast})=(D,ND)\stackrel{\subset}{\rightarrow}([D],[ND])=(\Psi_{3}(\sigma_{-}), \Psi_{3}^{ \ast}(\sigma_{+}))=([\ee^{\ii\theta }],[\ee^{\ii(-2\pi+\gamma )}]) ,\pi<\theta<\frac{3\pi}{2},\pi<\gamma< \frac{3\pi}{2}\nonumber
\end{equation}

\textbf{\em State 4.} induced from coupling between                                                
\begin{equation} 
\Psi_{4}^{(\vec{\mathcal{  S}},\vec{\mathcal{D}})}([s_{i}],[s_{j}]^{\ast})=(D,ND^{\ast})\stackrel{\subset}{\rightarrow}([D],[ND^{\ast}])=(\Psi_{4}(f_{+}), \Psi_{4}^{ \ast}(\sigma_{-}))=([\ee^{\ii\theta }],[\ee^{\ii(-2\pi+\gamma )}]) ,\pi<\theta<\frac{3\pi}{2},\frac{\pi}{2}<\gamma< \pi\nonumber
\end{equation}

\textbf{\em State 5.} induced from coupling between  
\begin{equation}                             
\Psi_{5}^{(\vec{\mathcal{S}},\vec{\mathcal{D}})}([s_{i}],[s_{j}]^{\ast})=(S^{\ast},NS)
\stackrel{\subset}{\rightarrow}
([U^{\ast}],[NU])=(\Psi_{5}(\sigma_{-}), \Psi_{5}^{ \ast}(f_{+}))=([\ee^{\ii\theta }],[\ee^{\ii(-2\pi+\gamma )}]) ,\frac{3\pi}{2}<\theta<2\pi,0<\gamma< \frac{\pi}{2}\nonumber
\end{equation}

\textbf{\em State 6.} induced  from coupling between                          
\begin{equation}
\Psi_{6}^{(\vec{\mathcal{S}},\vec{\mathcal{D}})}([s_{i}],[s_{j}]^{\ast})=(S^{\ast},NS^{\ast})
\stackrel{\subset}{\rightarrow}([U^{\ast}],[NU^{\ast}]) =(\Psi_{6}(f_{+}), \Psi_{6}^{ \ast}(f_{-}))=([\ee^{\ii\theta }],[\ee^{\ii(-2\pi+\gamma )}]) ,\frac{3\pi}{2}<\theta<2\pi,\frac{3\pi}{2}<\gamma< 2\pi\nonumber
\end{equation}

\textbf{\em State 7.} induced from coupling between                              
\begin{equation}
\Psi_{7}^{(\vec{\mathcal{S}},\vec{\mathcal{D}})}([s_{i}],[s_{j}]^{\ast})=(D^{\ast},ND)\stackrel{\subset}{\rightarrow}  ([D^{\ast}],[ND])=(\Psi_{7}(f_{-}), \Psi_{7}^{ \ast}(\sigma_{+}))=([\ee^{\ii\theta }],[\ee^{\ii(-2\pi+\gamma )}]) ,\frac{\pi}{2}<\theta<\pi,\pi<\gamma< \frac{3\pi}{2}\nonumber
\end{equation}

\textbf{\em State 8.} induced from coupling between 
\begin{equation}
\Psi_{8}^{(\vec{\mathcal{S}},\vec{\mathcal{D}})}([s_{i}],[s_{j}]^{\ast})=(D^{\ast},ND^{\ast})
\stackrel{\subset}{\rightarrow}([D^{\ast}],[ND^{\ast}])=(\Psi_{8}(f_{-}), \Psi_{8}^{ \ast}(f_{+})) =([\ee^{\ii\theta }],[\ee^{\ii(-2\pi+\gamma )}]) ,\frac{\pi}{2}<\theta<\pi,\frac{\pi}{2}<\gamma<\pi.\nonumber
\end{equation}

\end{Definition}

A market cyclic cohomology is induced from boundary operator between open string correlator
\begin{equation}
\partial C_{M}^{\phi}  (\Psi_{1}^{(\vec{\mathcal{S}},\vec{\mathcal{D}})},\Psi_{2}^{(\vec{\mathcal{S}},\vec{\mathcal{D}})},\cdots \Psi_{8}^{(\vec{\mathcal{S}},\vec{\mathcal{D}})})=\partial C_{M}^{\phi}  (\Psi_{2}^{(\vec{\mathcal{S}},\vec{\mathcal{D}})},\cdots \Psi_{8}^{(\vec{\mathcal{S}},\vec{\mathcal{D}})},\Psi_{1}^{(\vec{\mathcal{S}},\vec{\mathcal{D}})})
\end{equation}
with cyclic symmetry
\begin{equation}
C_{M}^{\phi}  (\Psi_{2}^{(\vec{\mathcal{S}},\vec{\mathcal{D}})},\cdots \Psi_{M}^{(\vec{\mathcal{S}},\vec{\mathcal{D}})},\Psi_{1}^{(\vec{\mathcal{S}},\vec{\mathcal{D}})})=(1)^{M-1+\Psi_{1}^{(\vec{\mathcal{S}},\vec{\mathcal{D}})}(\Psi_{2}^{(\vec{\mathcal{S}},\vec{\mathcal{D}})}+\cdots +\Psi_{M}^{(\vec{\mathcal{S}},\vec{\mathcal{D}})}) }  C_{M}^{\phi}  (\Psi_{1}^{(\vec{\mathcal{S}},\vec{\mathcal{D}})},\Psi_{2}^{(\vec{\mathcal{S}},\vec{\mathcal{D}})},\cdots \Psi_{M}^{(\vec{\mathcal{S}},\vec{\mathcal{D}})})
\end{equation}

\section{Methodology}\label{sec:method}
 
\subsection{Empirical analysis of financial tensor network for non-equilibrium state}
 
\begin{figure}[!t]
	\centering
	\includegraphics[height=0.7\textheight]{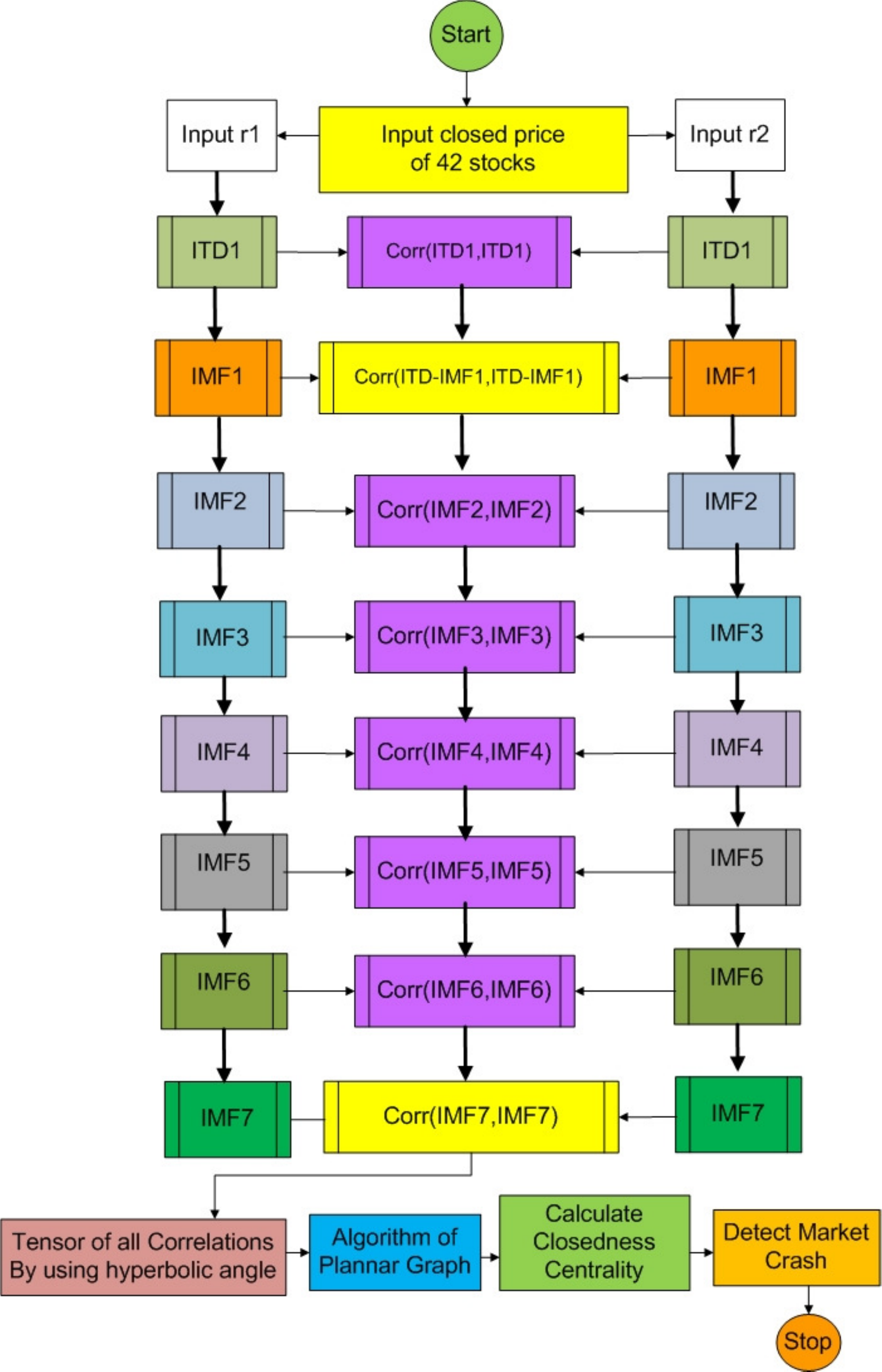}
	\caption{The flowchart of empirical analysis of $\ICHAIN(1,n)$ of close price in financial tensor network with closeness centrality. The algorithm in this flowchart use to detect market crash state in year 2008 from time series of close price of $42$ stocks in 2000 days. The output of this algorithm is a financial tensor network as planar graph \cite{network8} and a time series of closeness centrality of planar graph \label{flow_chart}.}
\end{figure}

If we consider two time series of close price of stocks $r_{i}(t)$, $r_{j}(t)$ we can break a mirror symmetry of financial market by Wigner ray $\m{SU}(2)$ transformation act on scalar product (see Def.~\ref{def:w}). In Euclidean geometry we measure a 2-form of angle, price $r_{j}$ by typical correlation formula. We visualize the process of financial tensor network \cite{tensor} by empirical analysis of a skeleton of time series data, i.~e., $\ICHAIN(1,n)$ (see \cite{kabin1}), with hyperbolic angle over financial tensor network \cite{tensor1}. The result is powerful to detect non-orientation state of market crash by closeness centrality measure over planar graph \cite{network8} over distance of partial correlation matrix \cite{kenett}. The schema of our methodology is presented in Fig.~\ref{flow_chart}.
 
Let $V_{\mathbb{R}}=:V_{\mathbb{R}}(M)$ be a vector of real value as a function of ground field of underlying financial market network $M$. We can extend ground field to $\mathbb{C}$, $\mathbb{H}$ and $\mathbb{H}P^{1}$. We define their dual vector fields of $42$ stock as eigenvector of vector space $V$  by $V=<<v_{1},v_{2},\cdots ,v_{42}>>$ (the full list of $42$  stocks is shown in Tab.~\ref{chinese}). We consider
\begin{equation}
V\simeq T_{p}M,\;\; \m{dim} V=42 ,\;\; V_{\mathbb{R}}^{\ast}:V\rightarrow \mathbb{R},
\end{equation}
where $T_{p}M$ is a tangent of manifold $M$. In behavior model we extend a field from $\mathbb{R}$ to $\mathbb{H}$ so we get one form of ground field with $8$ states by 
\begin{equation}
V_{\mathbb{H}P^{1}}^{\ast}:V\rightarrow \mathbb{H}P^{1},\;\; V_{\mathbb{H}P^{1}}^{\ast}=\wedge^{1} V_{\mathbb{H}p^{1}}.
\end{equation}
Let $r_{i},r_{j},i,j=1,cdots, 42$. The correlation matrix in financial network classically induces from linear map between vector space of market. Let $r_{1}$, $r_{2}$ be returns of stocks $stock_{1}$, $stock_{2}$ in financial market $stock_{1}, stock_{2}\in M$. We make an assumption that there exist dual market $M^{\ast}$, hidden financial market with hidden stocks $stock_{1}^{\ast}, stock_{2}^{\ast}\in M^{\ast}$.  Let $r_{12}$ be correlation of return of $stock_{1} $ and $stock_{2},r_{12}=\m{Corr}(r_{1},r_{2})$, it induces a hidden correlation between hidden stocks $r_{12}^{\ast}=\m{Corr}(r_{1},r_{2})^{\ast}$ and the correlation matrix between two markets $M\otimes M^{\ast}=\m{Corr}(r_{1},r_{2})\otimes \m{Corr}(r_{1},r_{2})^{\ast}.$ The precised geometrical definition  between two stocks in financial market $M$ is defined by 
\begin{equation}
\m{Corr}(M)=  \left[   \begin{array}{ccc}
&stock_{1}^{\ast}(t) &stock_{2}^{\ast}(t)\\
stock_{1}(t)&r_{11}&r_{12}\\
stock_{2}(t)&r_{21}&r_{22}\\
\end{array}\right ]\in \m{SO}(2)\simeq {S}^{1}\leftarrow \m{Spin}(2)\simeq S^{3},
\end{equation}
where $\simeq$ is a homotopy equivalent.
Because the value of correlation is from the interval $[-1,1]\subset \mathbb{R}$, the equilibrium property of market is coming from isometry property of the determinant of correlation matrix.
 
First we consider determinant of correlation matrix between two stocks for zero value,
\begin{equation}
\det\m{Corr}(M)=  \left|   \begin{array}{ccc}
&stock_{1}^{\ast}(t) &stock_{2}^{\ast}(t)\\
stock_{1}(t)&r_{11}&r_{12}\\
stock_{2}(t)&r_{21}&r_{22}\\
\end{array}\right |=0.
\end{equation}
Since $r_{11}=r_{22}=1$, $r_{12}=r_{21}$, we get
\begin{equation}
r_{12}r_{21}=1.
\end{equation}
If we generalized to  hermitian product of correlation in complex form, we get 
\begin{equation}
r_{12}r_{21}=<r_{12},r_{12}>=1.
\end{equation}
Thus, the isometry property of market induces an inertia frame of reference from the Hermitian property of correlation in complex conjugate $r_{12}^{\dagger}=r_{21}\in \mathbb{C}$. We have $r_{12}^{2}=1$ so $r_{12}=\pm 1$. In the state with $r_{12}=1$ the $stock_{1}$ is in the same direction of $stock_{2}$. For $r_{12}=-1$, the $stock_{1}$ is in the opposite direction of $stock_{2}$. These two states are the explicit states of financial market.
 
In complex projective space, we can extend the value of correlation to general form in Riemannian manifold $(M,g)$ by extend $r_{12},r_{21}\in \m{\mathbb{C}P}^{1}$, $r_{12}=r_{21}^{\dagger},$ where $\dagger$ is Hermitain product in quantum mechanics. We define metric tensor $g$ by using a differential 2-form over vector space $M$ of correlation between two stocks (correlation in complex projective space (Kähler manifold) in this sense).
 
The determinant of correlation matrix or a wedge product of column of correlation of all returns of stocks in stock market can be minus one and it induces a complex structure as a spin structure in principle bundle of time series data. We found that when $\det(\m{Corr}(M))^{2}=1$, where $\m{Corr}(M)$, $M\in \m{SL}(2,\mathbb{C})$ is a correlation matrix of financial network with Möbius map $M$ \cite{corr1,stock_stanley2}. This space of time series data is in equilibrium with two equilibrium points.
One is real and explicit form, the other is hidden and in complex structure like dark matter or invisible hand in economics concept. If $\det(\m{Corr}(M))=0$ a space of time series data is out of equilibrium and can induces market crash with more herding behavior of noise trader. This result is related to definition of log return and arbitrage opportunity in econophysics.
 
We can also compute a hyperbolic angle by using Hilbert–Huang transform of financial time series with the instantaneous frequency IMF$1$ of return $r_{i}$
\begin{equation}
\theta_{t}^{i}=\arg[\m{HHT}(r_{t}^{i})],\quad \theta_{t}^{j}=\arg[\m{HHT}(r_{t}^{j})]
\end{equation}
which induce an instantaneous correlation as instantaneous frequency by
\begin{equation}
\cos( \theta_{t})=\m{Corr}_{t}(r_{i},r_{j}).
\end{equation}
If we interchange between complex plane to hyperbolic plane by break a symmetry, the cosine function can be replaced by the hyperbolic cosine
\begin{equation}
\cosh( \Psi_{t})=\m{Corr}_{t}(r_{i},r_{j}).
\end{equation}
We have 
\begin{equation}
\theta_{t}=\arccos(\m{Corr}_{t}(r_{i},r_{j})), \quad
\Psi_{t}=\arccosh(\m{Corr}_{t}(r_{i},r_{j}))
\end{equation}
with help of
\begin{equation}
\frac{\ee^{\ii\theta}+\ee^{-\ii\theta}}{2}=\m{Corr}(r_{i},r_{j}).
\end{equation}
since $-\ii=\ee^{-\frac{\ii\pi}{2}}$ and $\ii=\ee^{\frac{\ii\pi}{2}}$, the Pauli matrix $\sigma_{x}:=f$ and the Wilson loop for time series data of Pauli matrix to be $\sigma=W(\sigma_{x})$ defined by
\begin{equation}
\frac{\ee^{\Psi\frac{\pi}{2}}-\ee^{-\Psi\frac{\pi}{2}}}{2}=\frac{1}{2}\det \left[   \begin{array}{cc}
0&\ee^{-\frac{\pi}{2}\Psi} \\
\ee^{\frac{\pi}{2}\Psi}&0\\
\end{array} 
\right ]        =\m{Corr}(r_{i},r_{j}).
\end{equation}

\subsection{Closeness centrality of planar graph in tensor network}
 
When market crashes the stocks in financial market will fall down in the same direction. Therefore the correlation between all stocks will be very high, it will induce approaching of all vertices to the others (on average). The closeness centrality is the most suitable to measure market crash of tensor financial network of planar graph. The time varying distance is defined by $d^{ij}_{t}=\sqrt{2(1-\m{Corr}(r_{i},r_{j}))}$. Let $G_{t}$ be random variable of graph with $G_{t}=(V_{t},E_{t})$, where $V_{t}$ is a time series  of vertex $V_{t} \in X_{t} $ of a 1-dim CW complex, i.~e., connected graph, of realization of Kolmogorov space for time series data $X_{t}$ and $E_{t}$ is time series of edge.
The time varying closeness centrality is defined by
\begin{equation}
C_{t}(k)=\frac{1}{\sum_{h\in G_{t}} d_{G_{t}}(h,k)}
\end{equation} 
Since the covering space of correlation matrix is a spin group of Pauli matrix. The map between them, we define in this work as modified Wilson loop for time series data. The algorithm to do minimum spanning tree \cite{minspan} and planar graph \cite{planar} is analogy to discrete dynamic programming approach to optimized an one dimensional CW complex of financial tensor network with $\theta$ as degree (sum of edge in vertex), as maximized parameter  in financial tensor network. That mean the special case of tensor field of correlation matrix, $\m{Corr}(M)\otimes \m{Corr}(M)^{\ast}$ can use the formular $\cosh(\m{Corr}(M)) := \m{Corr}^{\ast}((\m{Corr}(M)))$ and $\m{Corr}(\m{Corr}(M)) := \cos(\m{Corr}(M))$ for some special case of some type of tensor field. We do tensor in quantum information theory by using Kronecker product of matrix of correlation of $42$ stocks. We get $42\times 42$ result of matrix of correlation in tensor field. After that we use average correlation to find an average of matrix and plot into time series of correlation in tensor field. The tensor matrix of correlation can be compute by hand as followings,
\begin{align} 
&\left[ 
\begin{array}{cccc}
\m{Corr}(\mathrm{IMF}_{i}(r_{1}),\mathrm{IMF}_{i}(r_{1}))&\m{Corr}(\mathrm{IMF}_{i}(r_{1}),\mathrm{IMF}_{i}(r_{2}))&\cdots &\m{Corr}(\mathrm{IMF}_{i}(r_{1}),\mathrm{IMF}_{i}(r_{n}))\\
\m{Corr}(\mathrm{IMF}_{i}(r_{2}),\mathrm{IMF}_{i}(r_{1}))&\m{Corr}(\mathrm{IMF}_{i}(r_{2}),\mathrm{IMF}_{i}(r_{2}))&\cdots &\m{Corr}(\mathrm{IMF}_{i}(r_{2}),\mathrm{IMF}_{i}(r_{n}))\\
\cdots&\cdots &\cdots &\cdots \nn\\
\m{Corr}(\mathrm{IMF}_{i}(r_{n}),\mathrm{IMF}_{i}(r_{1}))&\m{Corr}(\mathrm{IMF}_{i}(r_{n}),\mathrm{IMF}_{i}(r_{2}))&\cdots &\m{Corr}(\mathrm{IMF}_{i}(r_{n}),\mathrm{IMF}_{i}(r_{n}))\\
\end{array}
\right ]_{42\times 42,\;i=1,\cdots n} \\
\otimes&
\left[ 
\begin{array}{cccc}
\m{Corr}(\mathrm{IMF}_{i}(r_{1}),\mathrm{IMF}_{i}(r_{1}))&\m{Corr}(\mathrm{IMF}_{i}(r_{1}),\mathrm{IMF}_{i}(r_{2}))&\cdots &\m{Corr}(\mathrm{IMF}_{i}(r_{1}),\mathrm{IMF}_{i}(r_{n}))\\
\m{Corr}(\mathrm{IMF}_{i}(r_{2}),\mathrm{IMF}_{i}(r_{1}))&\m{Corr}(\mathrm{IMF}_{i}(r_{2}),\mathrm{IMF}_{i}(r_{2}))&\cdots &\m{Corr}(\mathrm{IMF}_{i}(r_{2}),\mathrm{IMF}_{i}(r_{n}))\\
\cdots&\cdots &\cdots &\cdots\\
\m{Corr}(\mathrm{IMF}_{i}(r_{n}),\mathrm{IMF}_{i}(r_{1}))&\m{Corr}(\mathrm{IMF}_{i}(r_{n}),\mathrm{IMF}_{i}(r_{2}))&\cdots &\m{Corr}(\mathrm{IMF}_{i}(r_{n}),\mathrm{IMF}_{i}(r_{n}))\\
\end{array}
\right ]_{42\times 42,\;i=1,\cdots n} 
\end{align}
 
\begin{equation}
=\left[ 
\begin{array}{cccc}
\m{Corr}(\m{Corr}(c_{i}(r_{1}),c_{i}(r_{1})),\m{Corr}(c_{i}(r_{1}),c_{i}(r_{1})))  & \cdots &
\m{Corr}(\m{Corr}(c_{i}(r_{1}),c_{i}(r_{1})),\m{Corr}(c_{i}(r_{1}),c_{i}(r_{n})))  & \cdots\\
\m{Corr}(\m{Corr}( c_{i}(r_{1}), c_{i}(r_{1})),\m{Corr}(c_{i}(r_{2}),c_{i}(r_{1})))& \cdots &\m{Corr}(\m{Corr}(c_{i}(r_{1}),c_{i}(r_{1})),\m{Corr}(c_{i}(r_{2}),c_{i}(r_{n}))) & \cdots\\
\cdots&\cdots &\cdots&\cdots\\
\m{Corr}(\m{Corr}(c_{i}(r_{1}),c_{i}(r_{1})),\m{Corr}(c_{i}(r_{n}),c_{i}(r_{1})))  & \cdots &\m{Corr}(\m{Corr}(c_{i}(r_{1}),c_{i}(r_{1})),\m{Corr}(c_{i}(r_{n}),c_{i}(r_{n}))) & \cdots\\
\m{Corr}(\m{Corr}(c_{i}(r_{2}),c_{i}(r_{1})),\m{Corr}(c_{i}(r_{1}),c_{i}(r_{1})))  & \cdots
&\m{Corr}(\m{Corr}(c_{i}(r_{2}),c_{i}(r_{1})),\m{Corr}(c_{i}(r_{1}),c_{i}(r_{n}))) & \cdots\\
\m{Corr}(\m{Corr}(c_{i}(r_{2}),c_{i}(r_{1})),\m{Corr}(c_{i}(r_{2}),c_{i}(r_{1})))  & \cdots &\m{Corr}(\m{Corr}(c_{i}(r_{2}),c_{i}(r_{1})),\m{Corr}(c_{i}(r_{2}),c_{i}(r_{n}))) & \cdots\\
\cdots&\cdots &\cdots&\cdots\\
\m{Corr}(\m{Corr}(c_{i}(r_{2}),c_{i}(r_{1})),\m{Corr}(c_{i}(r_{n}),c_{i}(r_{1})))  & \cdots &\m{Corr}(\m{Corr}(c_{i}(r_{2}),c_{i}(r_{1})),\m{Corr}(c_{i}(r_{n}),c_{i}(r_{n}))) & \cdots\\
\cdots&\cdots &\cdots&\cdots\\
\end{array}
\right ]_{1764\times 1764}\nonumber
\end{equation}
 
\begin{equation}
=\left[ 
\begin{array}{cccc}
\cosh(\m{Corr}(c_{i}(r_{1}),c_{i}(r_{1})),\m{Corr}(c_{i}(r_{1}),c_{i}(r_{1})))  & \cdots &
\cosh(\m{Corr}(c_{i}(r_{1}),c_{i}(r_{1})),\m{Corr}(c_{i}(r_{1}),c_{i}(r_{n})))  & \cdots\\
\cosh(\m{Corr}(c_{i}(r_{1}),c_{i}(r_{1})),\m{Corr}(c_{i}(r_{2}),c_{i}(r_{1})))  & \cdots &\cosh(\m{Corr}(c_{i}(r_{1}),c_{i}(r_{1})),\m{Corr}(c_{i}(r_{2}),c_{i}(r_{n}))) & \cdots\\
\cdots&\cdots &\cdots &\cdots\\
\cosh(\m{Corr}(c_{i}(r_{1}),c_{i}(r_{1})),\m{Corr}(c_{i}(r_{n}),c_{i}(r_{1})))  & \cdots &\cosh(\m{Corr}(c_{i}(r_{1}),c_{i}(r_{1})),\m{Corr}(c_{i}(r_{n}),c_{i}(r_{n}))) & \cdots\\
\cosh(\m{Corr}(c_{i}(r_{2}),c_{i}(r_{1})),\m{Corr}(c_{i}(r_{1}),c_{i}(r_{1})))  & \cdots
&\cosh(\m{Corr}(c_{i}(r_{2}),c_{i}(r_{1})),\m{Corr}(c_{i}(r_{1}),c_{i}(r_{n}))) & \cdots\\
\cosh(\m{Corr}(c_{i}(r_{2}),c_{i}(r_{1})),\m{Corr}(c_{i}(r_{2}),c_{i}(r_{1})))  & \cdots
&\cosh(\m{Corr}(c_{i}(r_{2}),c_{i}(r_{1})),\m{Corr}(c_{i}(r_{2}),c_{i}(r_{n}))) & \cdots\\
\cdots&\cdots &\cdots &\cdots\\
\cosh(\m{Corr}(c_{i}(r_{2}),c_{i}(r_{1})),\m{Corr}(c_{i}(r_{n}),c_{i}(r_{1})))  & \cdots &\cosh(\m{Corr}(c_{i}(r_{2}),c_{i}(r_{1})),\m{Corr}(c_{i}(r_{n}),c_{i}(r_{n}))) & \cdots\\
\cdots&\cdots &\cdots &\cdots\\
\end{array} 
\right ]_{1764\times 1764} \nn
\end{equation}

\section{Results}\label{sec:results}

\begin{figure}[!t]
	\centering
	\includegraphics[width=0.8\textwidth]{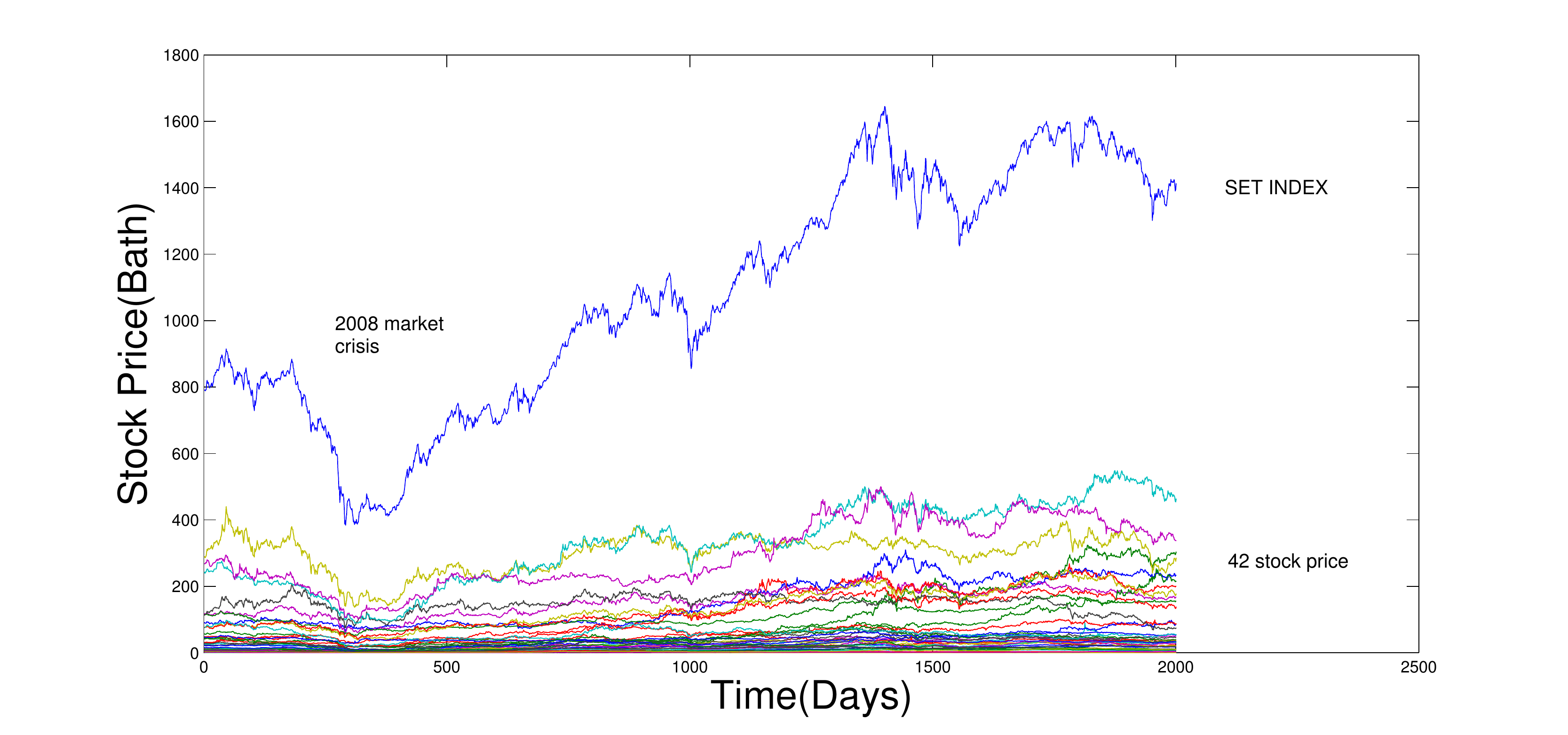} 
	\caption{The plot of close prices for 42 stocks of SET50 Index Futures during $2000$ days including 2008 market crash. The crash date is located under 272 point of the plot with all 42 stocks falling in the same direction down.\label{stock}}
\end{figure}

We used close prices of $42$ stocks in the Thai stock market SET50 Index Futures (Fig.~\ref{stock}), starting from 23/08/2007 to 29/10/2015 with $2000$ data points.
The results show the calculations for 2000 days interval with window size of 20 days. September 16th, 2008 as the crash point of financial market is located at date number 272. The details of all $42$ stocks are listed in Table~\ref{chinese} ($8$ stocks of SET50 Index Futures need to be cut out since the time duration of prices is insufficient for the calculation, the prices of other $42$ stocks pass through 2008 financial market crisis \cite{network}).

\begin{figure}[!t]
\centering
\includegraphics[width=0.49\textwidth]{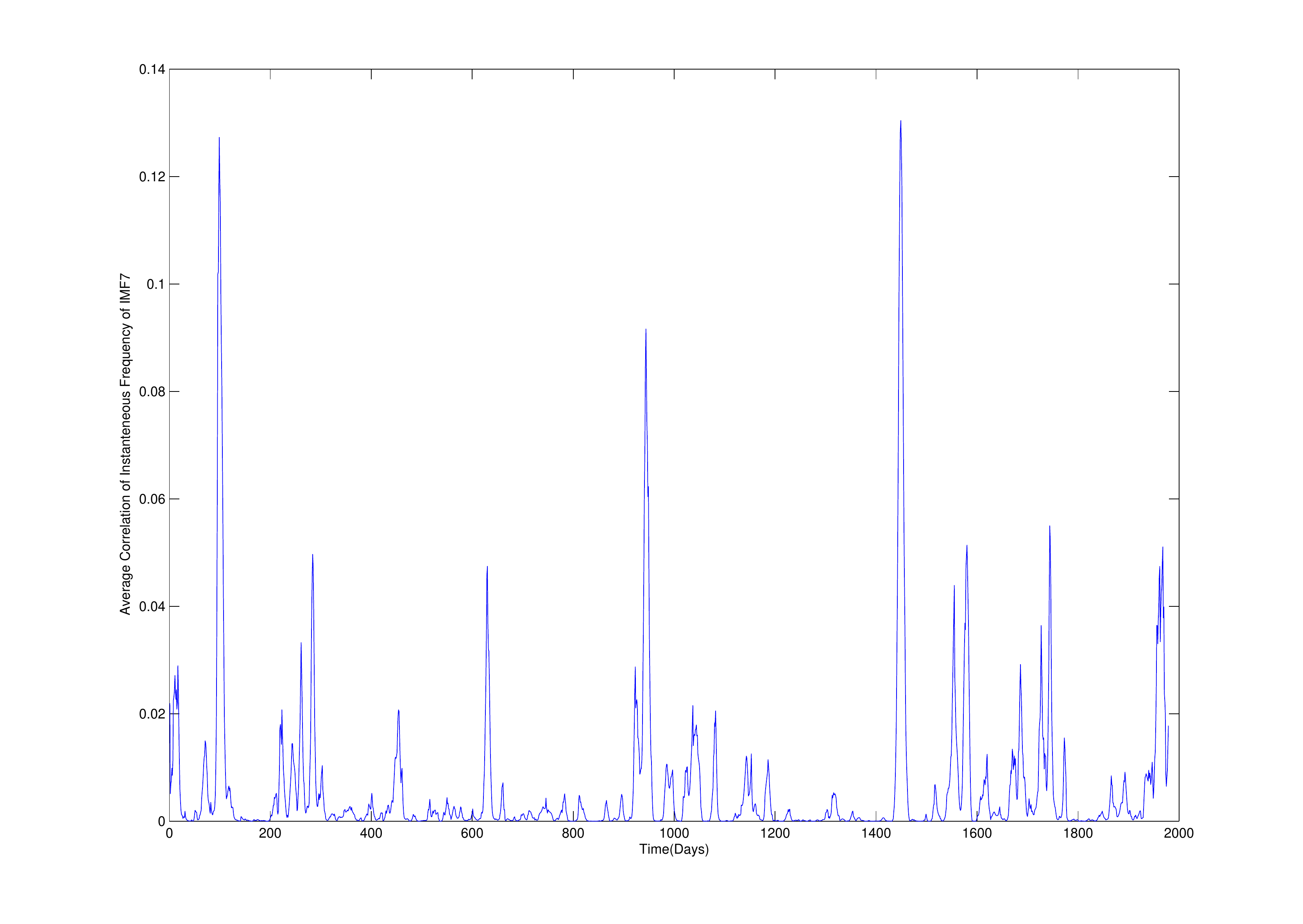}
\includegraphics[width=0.49\textwidth]{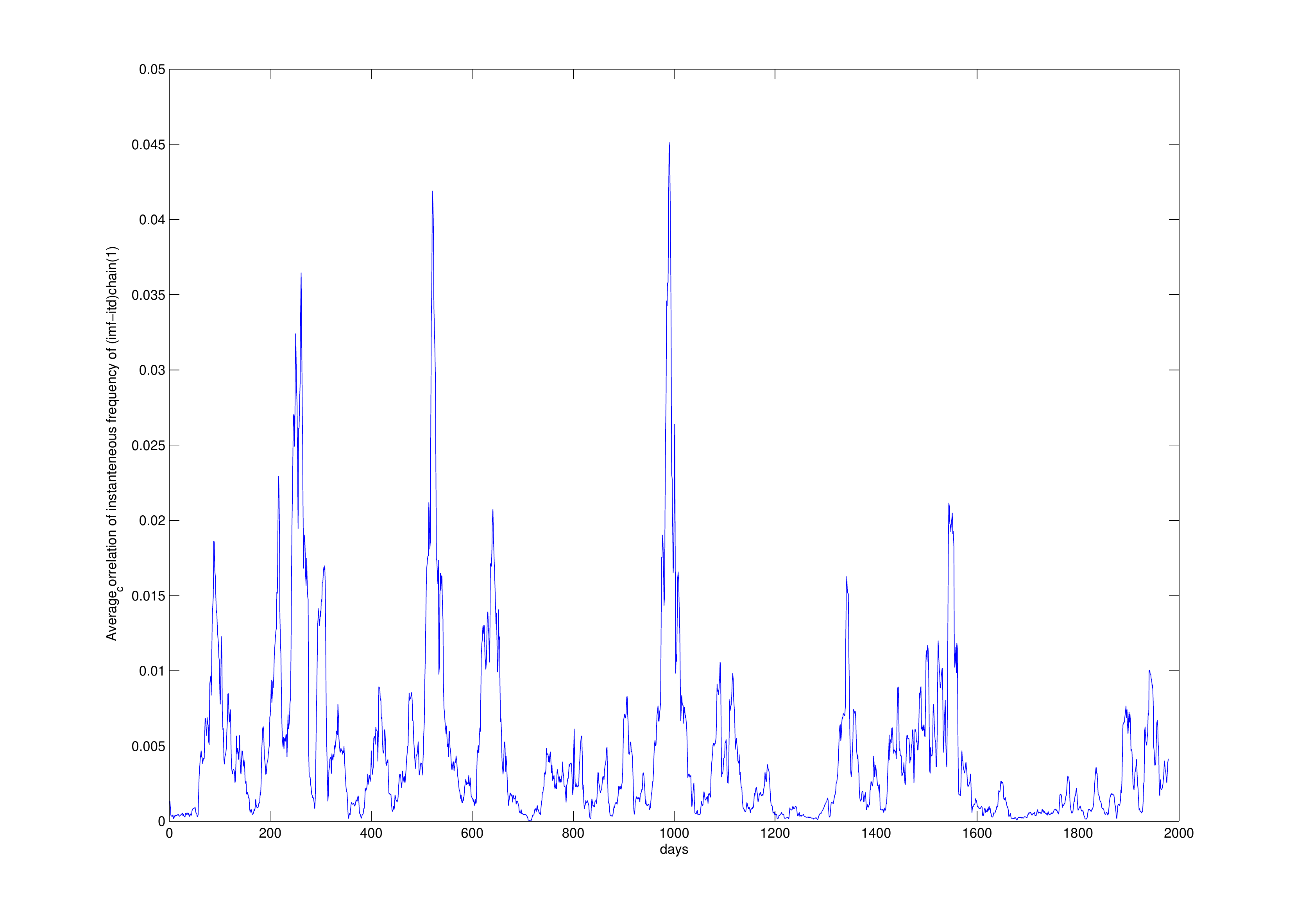}
\caption{The plot on left shows the average correlation of instantaneous frequency of IMF$7$. The right plot shows the average correlation of instantaneous frequency of $\ICHAIN(1,n)$ of all 42 stocks in 2000 days interval with window size of 20 days. The peak of 2008 market crash is located at the date $272$. \label{average_corr}}
\end{figure}

In Fig.~\ref{average_corr} we show the average correlation of instantaneous frequency IMF$7$ compared with the average correlation of instantaneous frequency of $\ICHAIN(1,n)$ for all $42$ stocks together. The peak of market crash at $272$ is relatively high but not highest for overall investigated period.

\begin{figure}[!t]
\centering
\includegraphics[width=0.49\textwidth]{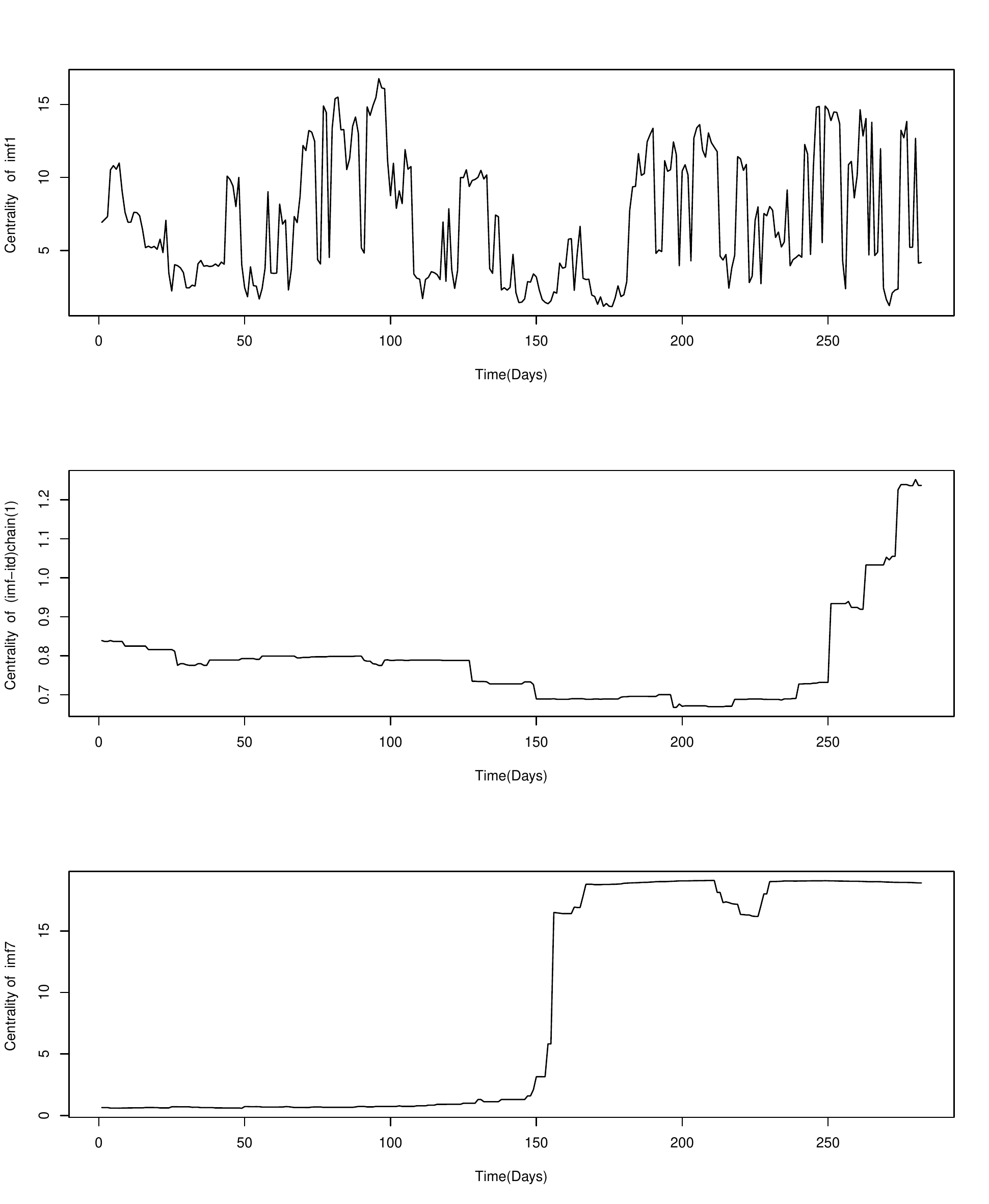}
\caption{The plots show a closeness centrality of IMF$1$ in above panel, $\ICHAIN(1,n)$ in the middle and IMF$7$ in the lower panel. The right limit of time axes (the date $272$) represents the date of 2008 financial market crisis. \label{plot_crash}}
\end{figure}

Next we have computed the closeness centrality of IMF$1$, $\ICHAIN(1,n)$ and IMF$7$ which are displayed in Fig.~\ref{plot_crash}. The end of time axes represents the market crisis date. One can see that the $\ICHAIN(1,n)$ of $42$ stocks has a very high peak and clearer result than IMF$1$ and IMF$7$.

\begin{figure}[!t]
\centering
\includegraphics[width=0.8\textwidth]{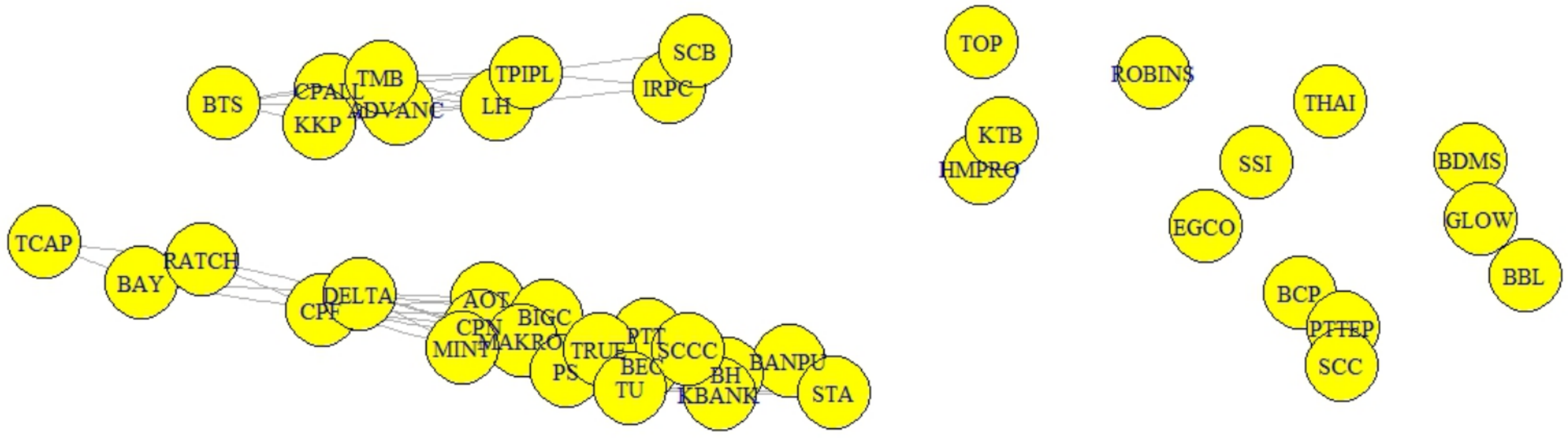}\\
\includegraphics[width=0.8\textwidth]{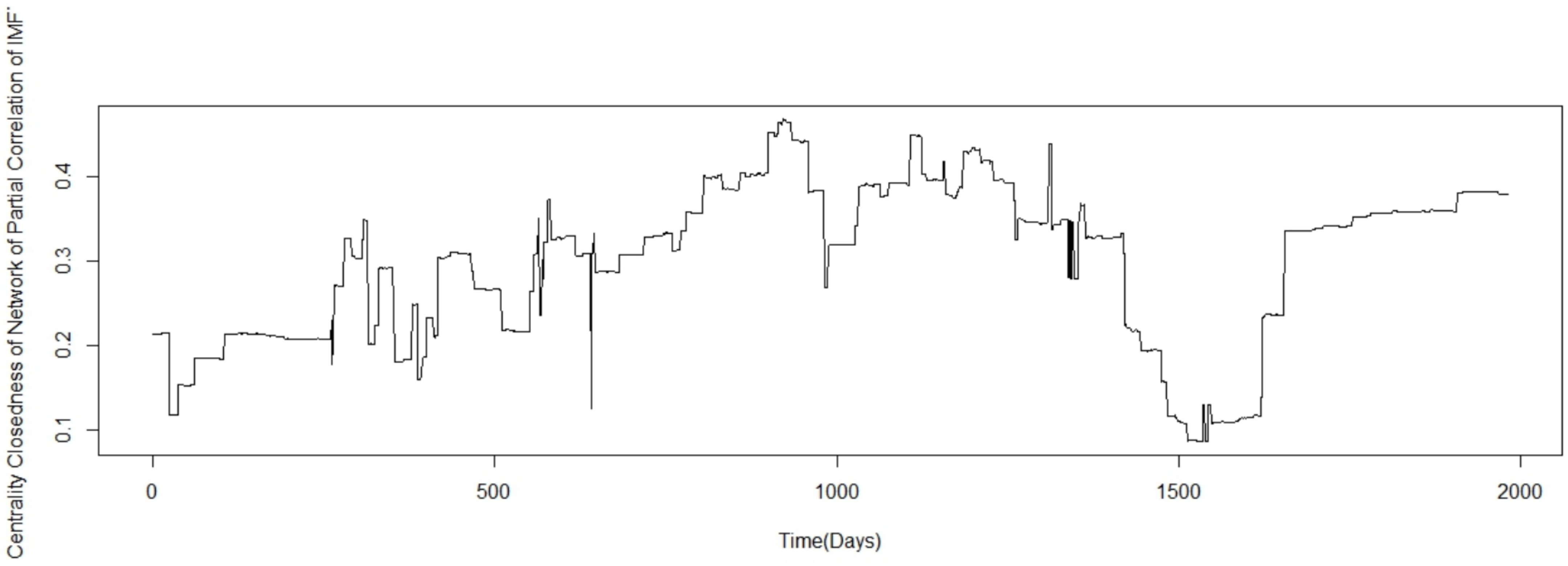}
\caption{The plot of closeness centrality of algorithm for hyperbolic cosine of $\ICHAIN(1,n)$ for $42$ stocks over the period of 2000 days with window size of 20 days. A tensor network of 42 stocks at the date 2000 is depicted above the plot.\label{central_imf_itd}}
\end{figure}

We have constructed the tensor network of tensor correlation to find the population of noise trader in financial market as demonstrated in our theory. We have used the hyperbolic cosine of $\ICHAIN(1,n)$. The result of closeness centrality together with graphical representation of tensor network of $42$ stocks of time series at the date 2000 is shown in Fig.~\ref{central_imf_itd}. One can see that the value of closeness centrality of 2008 market crash at the date 272 is significantly higher in comparison with neighborhood values. The figure also demonstrates the network separation into two clusters and some isolated nodes. The tensor network in our analysis is varying over time period in real time. When market crashes the network changes its structure topology and stocks are closer to each other, they are forming group.

\begin{figure}[!t]
\centering
\includegraphics[width=0.8\textwidth]{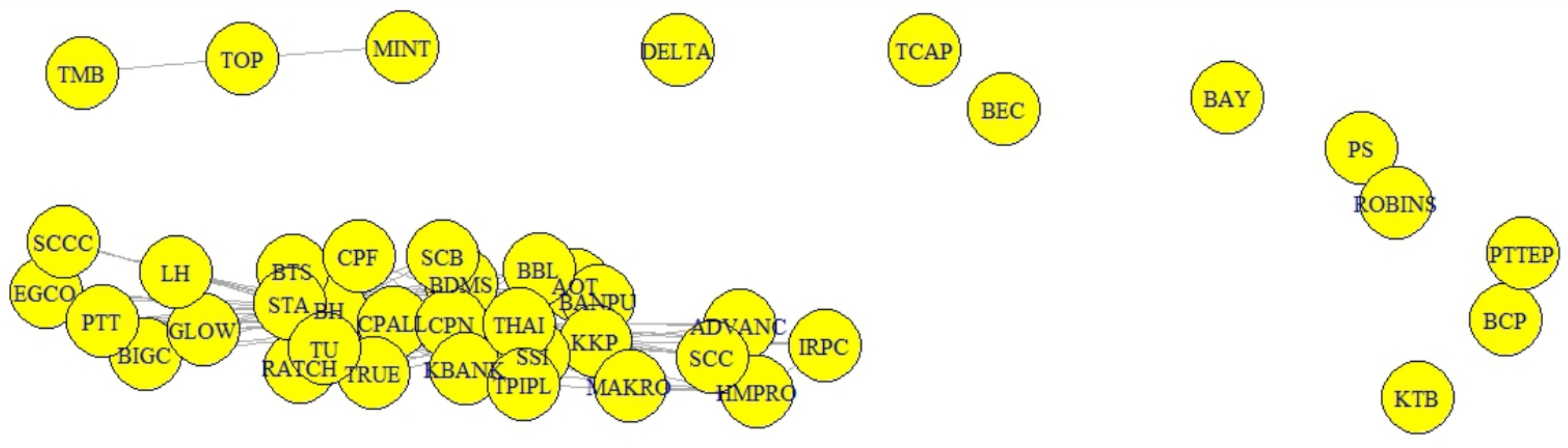}\\
\includegraphics[width=0.8\textwidth]{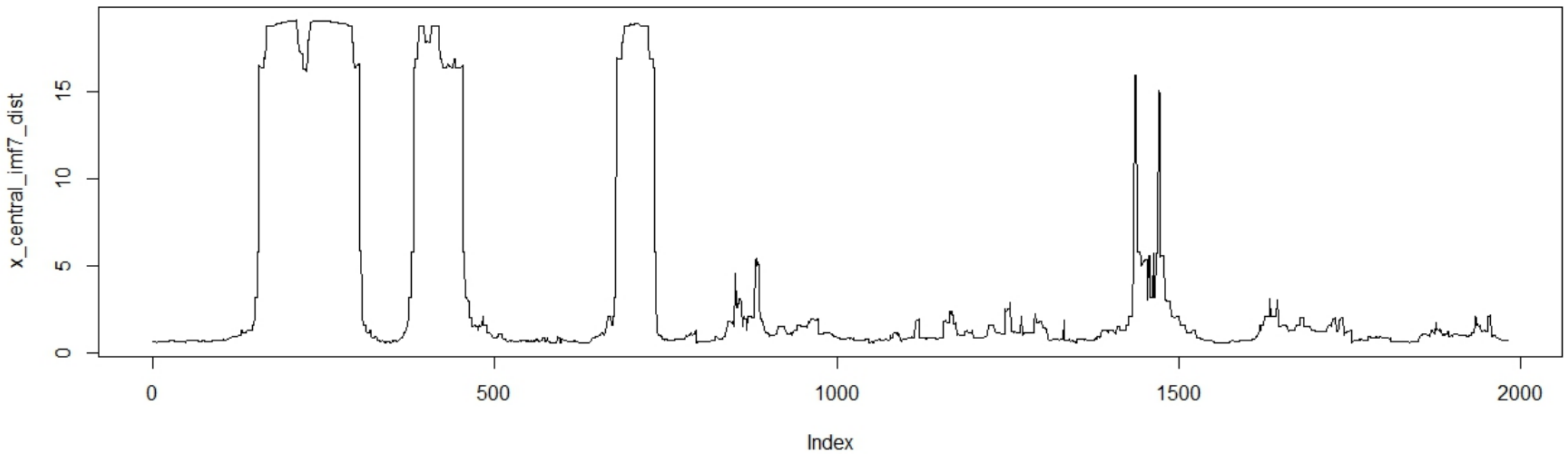}
\caption{The plot of closeness centrality of algorithm for IMF$7$ for $42$ stocks over the period of 2000 days with windows size of 20 days. The first peak is a capture of 2008 market crash state. A tensor network of all stocks at the date 2000 is depicted above the plot, it demonstrates the network separation into two clusters and some isolated nodes.\label{central_dis_imf7}}
\end{figure}
 
We have computed tensor networks from IMF$1$ to IMF$7$ to select which IMF is the best tool for a detection of 2008 market crash. We have found that the plot for closeness centrality of algorithm for IMF$7$ is the best tool to detect market crash, as is shown in Fig.~\ref{central_dis_imf7}. The first peak of closeness centrality is very thick with wider band than other peaks in the plot, it is a capture of 2008 market crash state. The tensor network above the plot demonstrates the situation at the date 2000, it is separated into two clusters and isolated nodes.

\section{Discussion and Conclusion}\label{sec:discussion}
    
Recently, econophysicists introduced a partial correlation matrix approach to analyze market structure empirically \cite{kenett}. The first and second eigen vectors of correlation matrix of stocks underlying financial market are denoted as eigen vectors of market in eigenportfolio model of market \cite{arbitrage}. Using this mathematical construction, the existence of a cohomology theory for financial market was taken into account in this work to be justified as the first cohomology theory in econophysics. The cohomology theory can be used to deform Kolmogorov space of time series data.

The meaning of correlation matrix of assets in financial market is deeply related to a risk analysis of portfolio management \cite{port} so called time-varying beta risk \cite{beta}. The geometric structure of correlation matrix is in a relation to the  Killing vector field and the covariant derivative of geometrical property of financial market.

The evolution of average edges of all 42 stocks underlying market is very high in the market crash state. The cohomology theory is used to construct 1-dim CW complex as planar graph in financial tensor network and to detect the topological defect in the market (the market crash state).

In this work, we have explained the existence of cohomology group over Kolmogorov space of time series data.  The general equilibrium in financial market exists when the sequence of market cocycle is an exact sequence. In the case of a complex projective space or Kähler manifold of market, the equilibrium point is in non-orientation state. 

We have developed a new theory so called cohomology theory for financial market with new series of definition of three types of market potential field with orientation and non-orientation state of entanglement state in modified Wilson loop of time series data to detect the entanglement state of market crash for Index Futures market. 
All Index Futures have underlying asset, for example SET50 Index Futures have $42$ selected stocks from $42$ underlying companies which are registered in Thailand stock market (SET) as based space and with SET50 Index Futures prices as tangent space. We have used the closeness centrality of a tensor field of partial correlation and the planar graph of Hilbert-Huang transform with the hyperbolic spectrum of instantaneous frequency as the main tools for a detection of market crash over time series data. The result of analysis shows that the closeness centrality of hyperbolic spectrum of IMF$7$ and $\ICHAIN(1,n)$ can be also used to detect the next market crashes.

 
\section*{Acknowledgments}
K. Kanjamapornkul is supported by Scholarship from the 100th Anniversary Chulalongkorn University Fund for Doctoral Scholarship. This research is supported by 90th Anniversary of Chulalongkorn  University, Rachadapisek Sompote Fund. The work was partly supported by VEGA Grant No. 2/0009/16 and APVV-0463-12. R. Pinčák would like to thank the TH division at CERN for hospitality. We would like to express our gratitude to Librade (\url{www.librade.com}) for providing access to their flexible platform, team and community. Their professional insights have been extremely helpful during the development, simulation and verification of the algorithms.

\bibliographystyle{unsrt} 
\bibliography{cohomology}

\begin{table}[!ht]
\renewcommand{\arraystretch}{1.3}
\caption{Table shows $42$ stocks in SET50 Index Futures of Thai stock market.}
\label{chinese}
\centering
\begin{tabular}{lllll}
\hline\hline
Rank &SET Symbol &Securities Name &Industry Group &Sector\\
\hline
1 &ADVANC &ADVANCED INFO SERVICE   &Technology &Information   Technology\\
2 &AOT &AIRPORTS OF THAILAND   &Services &Transportation  \\
3 &BANPU &BANPU   &Resources &Energy  \\
4 &BAY &BANK OF AYUDHYA   &Financials &Banking\\
5 &BBL &BANGKOK BANK   &Financials &Banking\\
6 &BCP &THE BANGCHAK PETROLEUM   &Resources &Energy\\  
7 &BEC &BEC WORLD   &Services &Media \\
8 &BDMS &BANGKOK DUSIT MEDICAL SERVICES   &Services &Health Care Services\\
9 &BH &BUMRUNGRAD HOSPITAL   &Services &Health Care Services\\
10 &BIGC &BIG C SUPERCENTER   &Services &Commerce\\
11 &BTS &BTS GROUP HOLDINGS   &Services &Transportation\\  
12 &CPALL &CP ALL   &Services &Commerce\\
13 &CPF &CHAROEN POKPHAND FOODS   &Agro Food Industry &Food and Beverage\\
14 &CPN &CENTRAL PATTANA   &Property  Construction &Property Development\\
15 &DELTA &DELTA ELECTRONICS (THAILAND)   &Technology &Electronic Components\\
16 &EGCO &ELECTRICITY GENERATING   &Resources &Energy  \\
17 &GLOW &GLOW ENERGY   &Resources &Energy  \\
18 &HMPRO &HOME PRODUCT CENTER   &Services &Commerce\\
19 &IRPC &IRPC   &Resources &Energy  \\
20 &KBANK &KASIKORNBANK   &Financials &Banking\\
21 &KKP &KIATNAKIN BANK   &Financials &Banking\\
22 &KTB &KRUNG THAI BANK   &Financials &Banking\\
23 &LH &LAND AND HOUSES   &Property  Construction &Property Development\\
24 &MAKRO &SIAM MAKRO   &Services &Commerce\\
25 &MINT &MINOR INTERNATIONAL   &Agro & Food and Beverage\\
26 &PS &PRUKSA REAL ESTATE   &Property   &Property Development\\
27 &PTT &PTT   &Resources &Energy  \\
28 &PTTEP &PTT EXPLORATION AND PRODUCTION   &Resources &Energy\\  
29 &RATCH &RATCHABURI ELECTRICITY                     &Resources &Energy\\  
   &      &                       GENERATING HOLDING  &          &      \\  
30 &ROBINS &ROBINSON DEPARTMENT STORE   &Services &Commerce\\
31 &SCB &THE SIAM COMMERCIAL BANK   &Financials &Banking\\
32 &SCC &THE SIAM CEMENT   &Property   &Property Development\\
33 &SCCC &SIAM CITY CEMENT   &Property   &Property Development\\
34 &SSI &SAHAVIRIYA STEEL INDUSTRIES   &Industrials &Steel\\
35 &STA &SRI TRANG AGRO-INDUSTRY   &Agro  & Food and Beverage\\
36 &TCAP &THANACHART CAPITAL   &Financials &Banking\\
37 &THAI &THAI AIRWAYS INTERNATIONAL   &Services &Transportation\\  
38 &TMB &TMB BANK   &Financials &Banking\\
39 &TOP &THAI OIL   &Resources &Energy  \\
40 &TPIPL &TPI POLENE   &Property  &Property Development\\ 
41 &TRUE &TRUE CORPORATION   &Technology &Information  \\
42 &TU &THAI UNION GROUP   &Agro  & Food and Beverage\\\hline\hline 
\end{tabular}
\end{table}

\end{document}